 \documentclass[11pt,a4paper]{article}
\usepackage{ifpdf} \usepackage{ae} \usepackage[T1]{fontenc}
\usepackage[ansinew]{inputenc} \usepackage{mathrsfs}
\usepackage{amsmath} \usepackage{amssymb} \usepackage{float}
\pdfoutput=1
\usepackage{amsmath,amssymb,amsfonts,a4wide,graphicx,bm,times,psfrag,wrapfig,sidecap}
\usepackage{cite}
\usepackage[colorlinks=true,linkcolor=black, citecolor=black,
urlcolor=black]{hyperref} 
\numberwithin{equation}{section}
\makeatletter \let\old@startsection=\@startsection
\renewcommand{\@startsection}[6]
{\old@startsection{#1}{#2}{#3}{#4}{#5}{#6\mathversion{bold}}}
\makeatother
\def\O{\Omega}

\def\defeq{\stackrel{\text{def}}=}
\newcommand\re[1]{({\ref{#1}})}
\def\be{\begin{eqnarray}} 
\def\ee{\end{eqnarray}}

\def\no{\nonumber}
\def\la{\label}
\def\l {\lambda}
\def\({\left(} \def\){\right)}
\def\<{\left\langle\,}
\def\>{\, \right\rangle}
\def\[{\left[}
\def\]{\right]}

\def\hf{ {\textstyle{1\over 2}} }

\def\CO{{\mathcal{ O} }}

\def\o{\omega}
\def\CC{{\mathcal C}}

\def\CN{{\mathcal N}}

\def\p{\partial}
\def\a{\alpha}
\def\b{\beta}

\def\s{\sigma}

\def\th{\theta}

\def\Tr{{\rm Tr}}
\def\Li{ \text{Li}_2}
\def\d{\delta}
\def\sla{\slash\!\! \!}

\newcommand{\caS}{{\mathscr S}}
\newcommand{\caA}{{\mathscr A}}

\def\zz{ { \{ z\} }}
\def\uu{ { \{ u\} }}
\def\vv{ { \{ v\} }}
\def\ww{ { \{ w\} }}
\def\zz  { { \{ \th \} }}

\def\1{ {(1)}}
\def\2{ {(2)}}
\def\3{ {(3)}}

\usepackage{bm}% bold math
\def\O{\Omega}
\def\defeq{\stackrel{\text{def}}=}
\def\ee{\end{eqnarray}}

\def\no{\nonumber} 
\def\la{\label} 
\def\l {\lambda}
\def\({\left(} 
\def\){\right)} 
\def\<{\left\langle\,} 
\def\>{\, \right\rangle} 
\def\[{\left[} 
\def\]{\right]} 

\def\hf{ {\textstyle{1\over 2}} }

\def\CO{{ \mathcal{ O} }}

\def\o{\omega}
\def\CC{ {\mathcal C}} 
 
\def\CN{{ \mathcal N}}

\def\p{\partial} 
\def\a{\alpha} 
\def\b{\beta}

\def\s{\sigma}

\def\th{\theta}

\def\Tr{{\rm Tr}}
\def\Li{ \text{Li}_2}
\def\d{\delta}
\def\sla{\slash\!\! \!}
\def\zz{ { { \bf  z} }}
\def\uu{ { {\bf u} }}
\def\vv{ { \bf v}}
\def\ww{ { \bf w }}
\def\thth{ { \bm {\th } }}
\def\vv { { \bf v  }}

\def\k{\kappa}

\def\k{\kappa}

\def\su3{ {^{su(3)}}}

\begin{document}

\begin{flushright}
 IPhT/t13/034
\end{flushright}

\vspace{1cm}
\setcounter{footnote}{0}

\begin{center}

 {\Large\bf A tree-level 3-point function in the ${ su(3)}$-sector of
planar $\mathcal{N}\!=\!4$ SYM}

\vspace{20mm} 

Omar Foda$^a$, 
Yunfeng Jiang $^b$
Ivan Kostov$^{b,}$\footnote{\it Associate member of the Institute for
Nuclear Research and Nuclear Energy, Bulgarian Academy of Sciences, 72
Tsarigradsko Chauss\'ee, 1784 Sofia, Bulgaria},
and  Didina Serban $^b$
 
 \vspace{15mm} 
 
 {$^a$\it Mathematics and Statistics,
Universityof Melbourne, \\
Parkville, Victoria 3010, Australia 
   \\[7mm]
 
{$^b$\it Institut de Physique Th\'eorique, CNRS-URA 2306 \\
	     C.E.A.-Saclay, \\
	     F-91191 Gif-sur-Yvette, France} \\[5mm]
}
 	    
\end{center}

\vskip9mm

\begin{abstract}
\noindent
We consider a particular case of the 3-point function of local 
single-trace operators in the scalar sector of planar $\mathcal{N} \!
= \!  4$ supersymmetric Yang-Mills, where two of the fields are 
 $su(3)$ type,  while the third one is  $su(2)$ type.
We show that this tree-level 3-point function can be expressed in terms of scalar
products of $su(3)$ Bethe vectors.  Moreover,  if the second level
Bethe roots of one of the $su(3)$ operators is trivial (set to
infinity), this 3-point function can be written in a determinant form.
Using the determinant representation, we evaluate the structure
constant in the semi-classical limit, when the number of roots goes to
infinity.
\end{abstract}

% 
%\begin{abstract}
%\noindent
%We classify the 3-point functions of local gauge-invariant
%single-trace operators in the scalar sector of planar $\mathcal{N} \!
%= \!  4$ supersymmetric Yang-Mills involving at least one $su(3)$
%operator.  In the case of two $su(3)$ and one $su(2)$ operators, the
%tree-level 3-point function can be expressed in terms of scalar
%products of $su(3)$ Bethe vectors.  Moreover, if the second level
%Bethe roots of one of the $su(3)$ operators is trivial (set to
%infinity), this 3-point function can be written in a determinant form.
%Using the determinant representation, we evaluate the structure
%constant in the semi-classical limit, when the number of roots goes to
%infinity.
%\end{abstract}

 %{\it  keywords: {Integrability, supersymmetric Yang-Mills}}
\begin{center}
{\tt\small
omar.foda at unimelb.edu.au, \\
yunfeng.jiang, ivan.kostov,
didina.serban at cea.fr
}
\end{center}

\vfill
\eject

\setcounter{footnote}{0}

\section{Introduction}

Computing $n$-point functions of local composite gauge-invariant
operators in $\mathcal{N} \!  = \!  4$ supersymmetric Yang-Mills
theory, SYM$_4$, is an important problem because these $n$-point
functions are among the basic objects on which the AdS/CFT 
correspondence can be tested
\footnote{ One of the early tests of the conjecture was to check 
that the tree-level $n$-point functions of the BPS operators 
coincide with those in supergravity 
\cite{LMRS,Rastelli-Cor}.}.
It is also a hard problem, even at tree-level, if only because 
of the combinatorial complexity of the operators involved.
However, developments over the past few years, starting with
\cite{Minahan:2002ve} and subsequent works, raise the hope that 
the methods of classical and quantum integrability can be used 
to solve this problem, at least in the planar limit.  
For a comprehensive review of integrability in SYM$_4$ and AdS/CFT, 
see \cite{Beisert-Rev} and references therein.  For shorter review, 
see \cite{Serban:2011aa}.

In this work, we focus on operators $\{\CO\}$ that are composed 
of fundamental fields in the scalar sector of SYM$_4$. Representing 
these fundamental fields as matrices in the adjoint representation 
of $su(N_c)$, $\{\CO\}$ are traces of products of 
$N_c\!  \times \!  N_c$ matrices. Further, in the planar limit that 
we are interested in, $N_c \rightarrow \infty$, multi-trace operators
are suppressed by factors of $1/N_c$, and one can take $\{\CO\}$ to 
be single-trace operators.

\subsection*{The weak-coupling limit} In weakly-coupled, perturbative
Yang-Mills theory, the computation of the $n$-point functions is a
well-defined problem.  Following Okuyama and Tseng
\cite{Okuyama-Tseng}, it is sufficient at tree-level to count all
possible planar sets of Wick contractions between the operators
involved.  Apart from normalization factors, the essential object in a
3-point function is the structure constant which, up to a
normalisation is a tri-linear form in the Hilbert space of states,
which we call {\it the cubic vertex}, in analogy with string field
theory.  In \cite{2004JHEP...09..032R}, Roiban and Volovich showed
that these $n$-point functions reduce to scalar products of spin-chain
states constructed using the algebraic Bethe Ansatz
\cite{Faddeev:1979gh}.  A systematic study in the case of three
operators that belong to (different) $su(2)$ sectors was presented by
Escobedo, Gromov, Sever and Vieira \cite{EGSV}.  The tree-level
correlation function of three $su(2)$ operators was expressed in
\cite{EGSV} in terms of scalar products of off-shell Bethe
states\footnote{If the magnon rapidities satisfy the Bethe equations,
the Bethe state is called {\it on-shell}, otherwise the Bethe state is
called {\it off-shell}.} of XXX spin-$\frac{1}{2}$ chains.  This
method is known as ``tailoring''.  Furthermore, it was shown in
\cite{Omar} that the 3-point function can be recast in terms of scalar
products of an off-shell state and an on-shell state and thereby can
be evaluated in determinant form.  We refer to this method as
``freezing''.

\subsection*{The semi-classical limit}
We are interested in computing the correlation functions of long
operators that are dual to semi-classical string states in AdS$_5$
$\times$ S$^5$.  The semi-classical (heavy) operators
$\{\CO_{\textit{sc}}\}$ are associated with classical solutions of the
string $\sigma$-model \cite{Beisert:2003xu, Kazakov:2004qf,
Beisert:2005bm}.  For a review see \cite{Schafer-Nameki:2010aa}.  In
spin-chain terms, the operators $\{\mathcal{O}_i\}$ are eigenstates of
the spin-chain Hamiltonian.  In other words, they are functions of
rapidity variables that satisfy Bethe equations.  For such operators,
the Bethe roots condense into several cuts (macroscopic Bethe strings)
in the complex rapidity plane \cite{Beisert:2003xu}.  The $n$-point
functions of semiclassical operators $\{ \CO_{\textit{sc}} \}$ are
particularly interesting, as they can be compared with the
corresponding correlation functions computed on the string theory
side.  Computing $n$-point functions of $\{ \CO_{\textit{sc}} \}$ in
the string theory was addressed in \cite{ Janik:2010aa,
Zarembo:2010ab, Buchbinder:2010ae, 2011arXiv1109.6262J,
2012PhRvD..85b6001B, 2011arXiv1106.0495K, 2011arXiv1110.3949K,
2012arXiv1205.6060K}.  However, the only case when the complete answer
is known is that of two heavy and one light operators
\cite{Zarembo:2010ab,Costa:2010rz,Roiban:2010aa}.  The same
configuration (heavy-heavy-light) was considered on the gauge theory
side by Escobedo {\it et al.} \cite{Escobedo:2011xw,
2011arXiv1107.5580C}, and in the $sl(2)$ sector by Georgiou
 \cite{2011JHEP...09..132G}.  They used a coherent state approximation for
the two heavy operators in the $su(3)$ sector.  Comparison with the
Frolov-Tseytlin limit \cite{Fr-Ts-lim} of the string theory result
\cite{Zarembo:2010ab,Costa:2010rz,Roiban:2010aa} showed a perfect
match.

The general case, when all three asymptotically-long operators are
non-BPS, the complete answer for the three-point function is known
only for weak coupling, and for special choice of the operators.  In
\cite{GSV}, Gromov, Sever and Vieira presented a thorough analysis of
the case of one BPS and two non-BPS heavy fields from the $su(2)$
sector.  In spin-chain terms, BPS operators are characterized by
trivial Bethe roots that are set to infinity.  The main result of
\cite{GSV} is an analytic contour integral derived from Korepin\rq{}s
sum expression for the scalar product of two off-shell states
\cite{korepin-DWBC}.  In \cite{SL,3pf-prl}, the determinant expression
obtained in \cite{Omar} was used to solve the problem in the general
case of three heavy non-BPS operators. 
 In \cite{2012arXiv1202.4103G,
Didina-Dunkl, 2012arXiv1205.5288G}, it was argued that this solution
gives the 3-point function at one  and two loops.   At one loop this 
conjecture was verified in  \cite{2012arXiv1202.4103G,2012arXiv1205.5288G}.

\subsection*{Outline of contents}
In Section {\bf \ref{sect:classification}}, we classify the 3-point
functions such that at least one operator is from the $su(3)$ sector.
In {\bf \ref{sect:su2.3pf=scpr}}, we recall the formulation of the
$su(2)$ 3-point function, including the cubic vertex, in determinant
form.  
{In {\bf \ref{sect:su3.3pf=scpr}}, we generalize the freezing method  
of ref.\cite{Omar} to the case where
two of the operators belong to $su(3)$ sectors while the third
belongs to an $su(2)$ sector.
%Then we use a result of
%\cite{Wheeler-SU3} to  write the  expression of the
%3-point function as a determinant.  
%, and the cubic vertex, obtained in \cite{Omar}, where
%all operators belong to scalar $su(2)$ sub-sectors.
%, to the case where
%two of the operators belong to $su(3)$ sub-sectors while the third
%belongs to an $su(2)$ sector. 
 Then we take one set of Bethe roots of one of
the $su(3)$ operators to be trivial (sent to infinity) and 
use the result of \cite{  Wheeler-SU3} to write the 3-point function in a determinant form.\footnote{A particular limit of the result of \cite{Wheeler-SU3} was previously obtained by Caetano and Vieira, see also ref.\cite{joaocaetano-scpr}.
}}
%, we need to take one set of Bethe roots of one of
%the $su(3)$ operators to be trivial (sent to infinity). 
 In {\bf
\ref{sect:dilog.su2}}, we recall the how the $su(2)$ 3-point function
was written in \cite{SL,3pf-prl,sz} in terms of certain functionals in
order to be able to compute its semi-classical limit.  In {\bf
\ref{sect:dilog.su3}}, we write the $su(3)$ 3-point function in terms
of the  quantities defined  in the previous section.  In {\bf
\ref{sect:semi-classical}}, we compute the semi-classical limit of the
$su(3)$ 3-point function of three non-BPS operators, under an
assumption that allows us to compute the semi-classical limit of the
norm of an $su(3)$ Bethe eigenstate.
%Section \ref{sect:conclusions} contains a number of remarks. 
Appendix \ref{appendix:nesting} contains a brief introduction 
to the nested coordinate Bethe Ansatz which is needed for the 
`tailoring' approach to the $su(3)$ 3-point function.
Appendix \ref{appendix:tailoring} includes details of the 
\lq tailoring\rq\ approach to the $su(3)$ 3-point functions. 
Appendix \ref{appendix:caA} discusses the properties of the 
functional forms that are needed to obtain the semi-classical 
limits.

\section{3-point functions with at least one $su(3)$
operator} \la{sect:classification}

\subsection{The structure constant  in $\CN=4$ SYM}

The 2-point and the 3-point functions are determined, up to
multiplicative factors, by conformal invariance,

\be \la{twopf} \langle \CO_i(x_i) \CO_j(x_j)\rangle =\ { L _i\ \CN_i \
\d_{i j} \over |x_1-x_2|^{2\Delta_i }}, \ee

\be \la{3ptfCFT} \langle \CO_1(x_1)\CO_2(x_2)\CO_3(x_3)\rangle =
{1\over N_c}\ {L_1L_2L_3\ \sqrt{ \CN_1 \CN_2 \CN_3 } \ \ \ C_{123}(\l)
\over |x_{12}|^{\Delta_1 +\Delta_2-\Delta_3} |x_{23}|^{\Delta_2
+\Delta_3-\Delta_1} |x_{31}|^{\Delta_3 +\Delta_1-\Delta_2} }, \ee

\noindent where $N_c$ is the number of colors, $\l$ is the {}\rq t
Hooft coupling and the three factors $\CN_i$ depend on the
normalization of the operators $\{ \CO_i(x_i)\}$.  The structure
constant $C_{123}(\l)$ does not depend on the normalization.

The factors $L_i$, equal to the number of fundamental operators in
$\CO_i$, account for the cyclic rotations of the trace operator
$\CO_i$.  The structure constant $C_{123}(\l) $ has the perturbative
expansion

 \be C_{123}(\l) = C^{(0)}_{123} + \l\ C^{(1)}_{123} + \dots \, .  \ee

To compute the tree-level structure constant $C_{123}^{(0)}$ using the
method of \cite{EGSV}, one needs the 1-loop wave functions.  At 1-loop
level, the operator $\CO_i$ is represented by a Bethe eigenstate with
energy $\Delta_i$.

 \bigskip
 
 \begin{figure}[h!]
         \centering
                 \begin{minipage}[t]{0.8\linewidth}
            \centering
            \includegraphics[width=7.0cm]{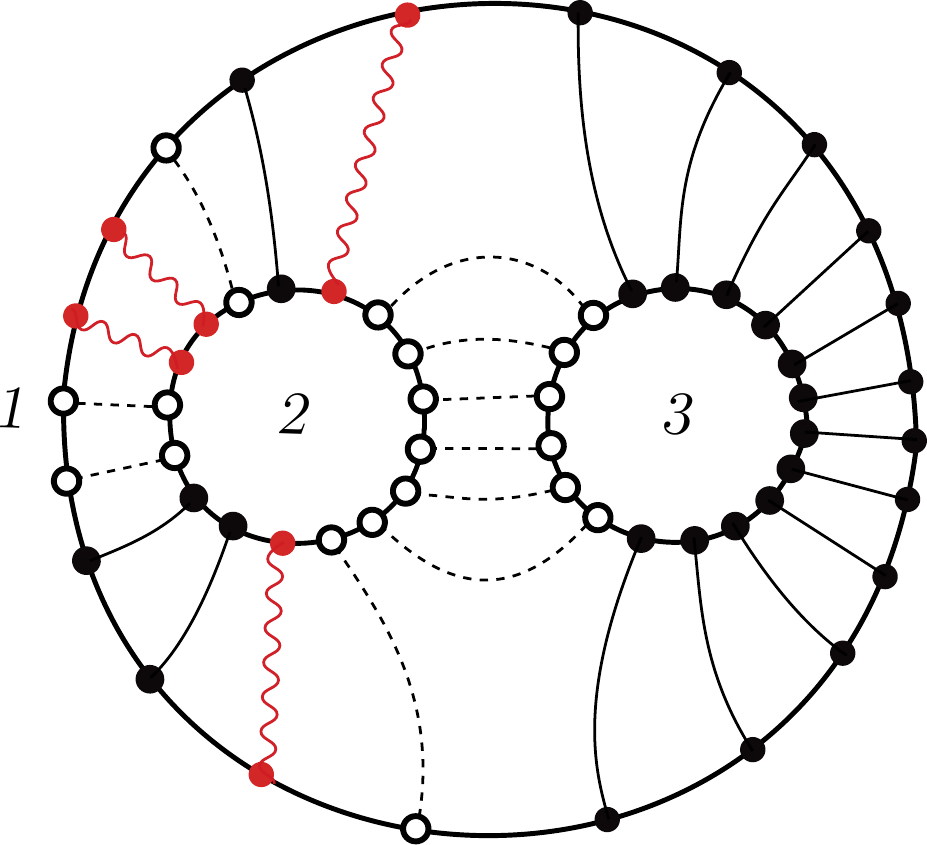}
 \caption{ \small Planar contractions contributing to the tree-level
 3-point function.  The contractions $\langle X\bar X\rangle, \langle
 Y\bar Y\rangle$ and $\langle Z\bar Z\rangle$ are represented
 respectively by black solid lines, blue solid lines and dashed
 lines.}
\label{fig:3pf-SU3}
\end{minipage}
\end{figure}

\bigskip

In a $su(2)$ sector, there is only one non-trivial configuration of
3-point functions.  In the presence of one or more operators from an
$su(3)$ sector, the structure of the 3-point functions becomes richer
and we need to classify the set of possible non-trivial configurations
of structure constants.  An example of a set of planar contractions is
given in Fig.  \ref{fig:3pf-SU3}.  The contractions $\<X\bar{X}\>$,
$\<Y\bar{Y}\>$ and $\<Z\bar{Z}\>$ are represented respectively by
solid lines, (red) wavy lines and dashed lines.

\begin{figure}[h!]
 %    \vskip 1cm
         %%----start of first figure----
	 \begin{minipage}[t]{0.35\linewidth}
            \centering
            \includegraphics[width=3.3 cm]{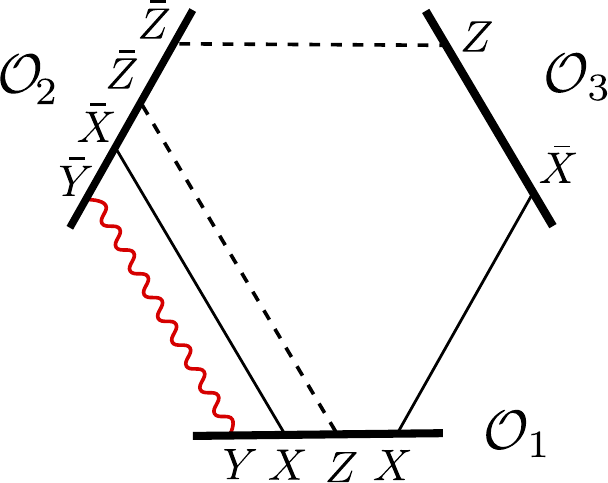}
\caption{ \small Schematic representation of the type-$\{2,3,3\}$
correlation function from Fig.  \ref{fig:3pf-SU3}.  }
  \label{fig:233a}
         \end{minipage}%
         \hspace{1.8cm}%
                \begin{minipage}[t]{0.45\linewidth}
            \centering
            \includegraphics[width=3.0cm]{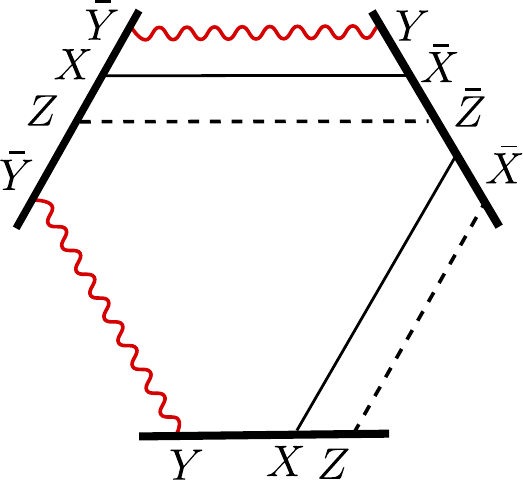}
           \hskip 0.3cm  \includegraphics[width=3.1cm]{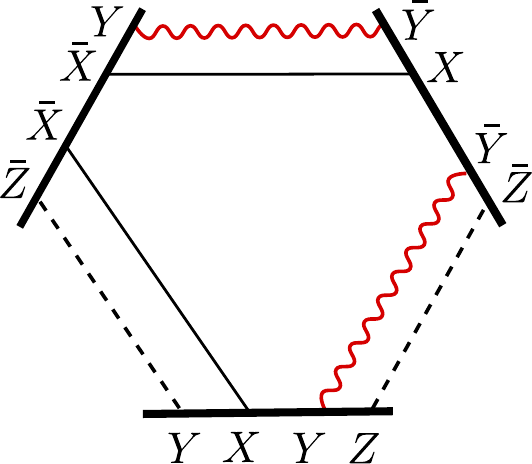}
	     \caption{ \small The two type-$\{3,3,3\}$ structure
	     constants.  }
\label{fig:333}
         \end{minipage}
 \vskip1cm
         %%----start of first figure----
	 \begin{minipage}[t]{0.50\linewidth}
        %    \centering
        \hskip 0.8cm     \includegraphics[width=3.1 cm]{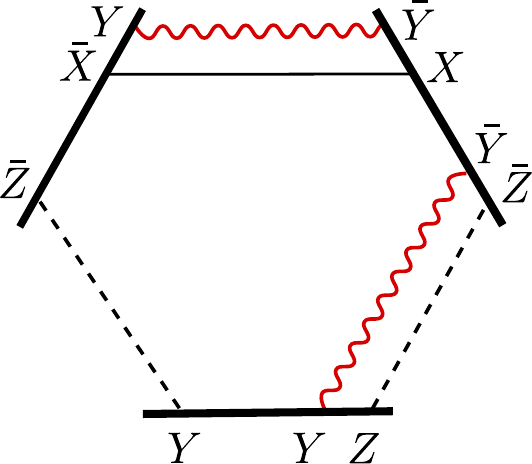}
         \hskip 0.5cm      \includegraphics[width=3.2 cm]{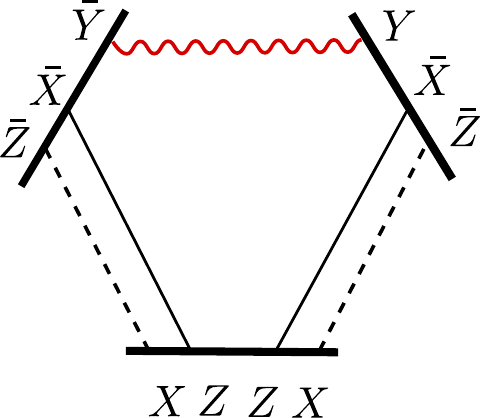}
\caption{ \small   The remaining two    type-$\{2,3,3\}$ structure constants. }
  \label{fig:233b}
         \end{minipage}%
         \hspace{1.2cm}%
                \begin{minipage}[t]{0.39\linewidth}
            \centering
            \includegraphics[width=3.1cm]{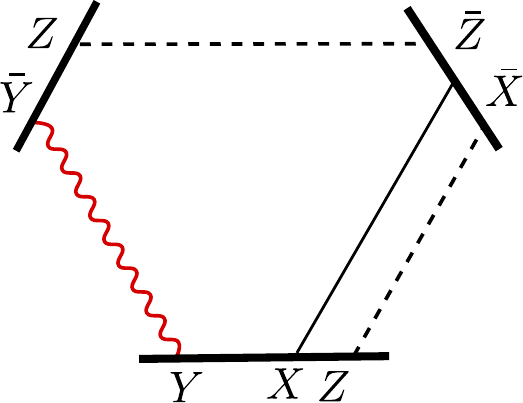}
	     \caption{ \small The   type-$\{2,2,3\}$ structure constant. }
\label{fig:223}
         \end{minipage}
% \vskip 1cm
      \end{figure}
%%%%%%%%%%%%%%%%%%%%%%%%%%%%%

Let us introduce some conventions.  There are several possible choices
of an $su(3)$ sector, which correspond to a choice of three distinct
complex scalar fields $X$, $Y$, $Z$, $\bar X$, $\bar Y$, $\bar Z$,
with pairs of mutually conjugated fields, like $Z$ and $\bar Z$,
excluded.  When only two types of non-conjugate scalar fields are
chosen, the composite operator belongs to an $su(2)$ sector.  If
$\mathcal{O}_1$, $\mathcal{O}_2$ and $\mathcal{O}_3$ belong to
$su(\alpha)$, $su(\beta)$ and $su(\gamma)$ sectors respectively, then
the corresponding 3-point function of type $\{\alpha, \beta,
\gamma\}$.  By permutation invariance, the order of $\alpha , \beta,
\gamma$ is irrelevant.

We represent the different classes of correlation functions
schematically by specifying the different types of Wick contractions
between pairs of operators.  For example, the correlation functions
corresponding to Fig.  \ref{fig:3pf-SU3} belong to the class
$\{2,3,3\}$ in Fig.  \ref{fig:233a}.  We call the operator at the
bottom $\mathcal{O}_1$, the one at right $\mathcal{O}_3$ and the one
at left $\mathcal{O}_2$.  Exchanging a scalar field and its complex
conjugate in all the operators does not change the value of the
structure constant.  This enables us to choose $\mathcal{O}_1$ such
that it contains only the scalar fields $X,Y$ and $Z$.  Since we are
interested in the large $N_c$ limit, only planar contractions are
retained.

We start by classifying the type-$\{3,3,3\}$ structure constants.  In
this case, there are two non-trivial inequivalent configurations, as
is shown in Fig.  \ref{fig:333}.  Deleting one line, that is, one type
of Wick contractions, from each of these two configurations, one
obtains type-$\{2,3,3\}$ structure constants.  There are three such
configurations, as shown in Fig.  \ref{fig:233a} and Fig.
\ref{fig:233b}.

Deleting one line from the configurations in Fig.  \ref{fig:233b}, one
obtains a type-$\{2,2,3\}$ or a type-$\{2,2,2\}$ 3-point functions.
The latter is a pure $su(2)$ 3-point function of the type studied in
\cite{EGSV, Omar, 3pf-prl, SL}.  There is one configuration of
type-$\{2,2,3\}$, as in Fig.  \ref{fig:223}.  To summarize, there are
six non-trivial types of 3-point functions with at least one $su(3)$
operator.

\subsection{Tailoring the tree-level structure 
constants}

Following   \cite{EGSV},  we construct  the structure 
constant   in three steps. 
{\bf 1.} We split the algebraic Bethe Ansatz representation of each 
spin chain into two: a left sub-chain and a right sub-chain. 
{\bf 2.} Considering each spin chain to be an in-state, we ``flip'' 
each left sub-chain from an in-state to an out-state. 
{\bf 3.} We take the scalar products of the left sub-chain state of 
$\mathcal{O}_i$ with the right sub-chain state of
$\mathcal{O}_{i+1\; {\rm mod}\; 3}$ $(i=1,2,3)$.  
Finally, we normalize the three external states. Further details on 
the tailoring procedure are in appendix \ref{appendix:tailoring}.  
We give our setup data in the table below.
\bigskip
\begin{center}
\begin{tabular}{|c|c|c|c|c|}
\hline
Operators & Length & Rapidities & No.  of Rapidities & Partitions of
Rapidities \\
\hline
$\mathcal{O}_1$ & $L_1$ & $\mathbf{u}_1,\mathbf{u}_2$ & \#
$\mathbf{u}_1$=$N_1$, \# $\mathbf{u}_2$=$M_1$ &
$\mathbf{u}'_1\cup\mathbf{u}''_1=\mathbf{u}_1$,
$\mathbf{u}'_2\cup\mathbf{u}''_2=\mathbf{u}_2$ \\
\hline
$\mathcal{O}_2$ & $L_2$ & $\mathbf{v}_1,\mathbf{v}_2$ & \#
$\mathbf{v}_1$=$N_2$, \# $\mathbf{v}_2$=$M_2$ &
$\mathbf{v}'_1\cup\mathbf{v}''_1=\mathbf{v}_1$,
$\mathbf{v}'_2\cup\mathbf{v}''_2=\mathbf{v}_2$ \\
\hline
$\mathcal{O}_3$ & $L_3$ & $\mathbf{w}_1,\mathbf{w}_2$ & \#
$\mathbf{w}_1$=$N_3$, \# $\mathbf{w}_2$=$M_3$ &
$\mathbf{w}'_1\cup\mathbf{w}''_1=\mathbf{w}_1$,
$\mathbf{w}'_2\cup\mathbf{w}''_2=\mathbf{w}_2$ \\
\hline
\end{tabular}
\end{center}
 \bigskip
The lengths of the left subchains are
\begin{align}
L_{13}&=\hf(L_1+L_3-L_2)\;,\\\nonumber
L_{12}&=\hf(L_1+L_2-L_3)\;,\\\nonumber L_{23}&=\hf(L_2+L_3-L_1)\;.
\end{align}
The structure constant reads
\begin{align}\label{c123}
 C_{123}^{(0)}=\sqrt{\frac{L_1L_2L_3}{\mathcal{N}_1\,
 \mathcal{N}_2\,\mathcal{N}_3}}\sum_{{\mathbf{u}}',
 {\mathbf{v}}',{\mathbf{w}}'}{\mathrm{H}}_F^{\mathbf{u}}\
 {\mathrm{H}}_F^{\mathbf{v}}\ {\mathrm{H}}_F^{\mathbf{w}} \
 \langle{\mathbf{u}}''^*|{\mathbf{v}}'\rangle \
 \langle{\mathbf{v}}''^*|{\mathbf{w}}'\rangle \
 \langle{\mathbf{w}}''^*|{\mathbf{u}}'\rangle\;,
\end{align}
where $\mathcal{N}_i$ are the norms of the Bethe states\footnote{In
this section all scalar products and norms are understood in the
Coordinate Bethe Ansatz normalization.}:
\begin{align}
\mathcal{N}_1=\langle{\mathbf{u}}|{\mathbf{u}}\rangle\;,\qquad
\mathcal{N}_2=\langle{\mathbf{v}}|{\mathbf{v}}\rangle\;,\qquad
\mathcal{N}_3=\langle{\mathbf{w}}|{\mathbf{w}}\rangle\;.
\end{align}
The $\mathrm{H}_F$ factors are given by
\begin{align}
\mathrm{H}^\mathbf{u}_F&=S_1 (\mathbf{u}''_1,\mathbf{u}_0) \ S_1
(\mathbf{u}''_2,\mathbf{u}_1)S_2^> (\mathbf{u}_1,\mathbf{u}''_1)S_2^>
(\mathbf{u}_2,\mathbf{u}''_2)\\\nonumber \mathrm{H}^\mathbf{v}_F&=S_1
(\mathbf{v}''_1,\mathbf{v}_0) \ S_1 (\mathbf{v}''_2,\mathbf{v}_1) \
S_2^> (\mathbf{v}_1,\mathbf{v}''_1) \ S_2^>
(\mathbf{v}_2,\mathbf{v}''_2) \\\nonumber \mathrm{H}^\mathbf{w}_F&=S_1
(\mathbf{w}''_1,\mathbf{w}_0) \ S_1 (\mathbf{w}''_2,\mathbf{w}_1) \
S_2^> (\mathbf{w}_1,\mathbf{w}''_1) \ S_2^>
(\mathbf{w}_2,\mathbf{w}''_2)\;.
\end{align}
with $\uu_0=\{0^{L_1+1}\},\; \vv_0=\{0^{L_2+1}\}$ and
$\ww_0=\{0^{L_3+1}\}$.  In the previous formula we used the following
notations: we denote the scattering factors as
\begin{align}
\label{s12main}
S_\s(u_{a,i},u_{b,j})=\frac{u_{a,i}-u_{b,j}
+\frac{i}{2}\s}{u_{a,i}-u_{b,j}-\frac{i}{2}\s}\;, \qquad \s=1,2\;,
\end{align}
and, given a function $F(x,y)$ and two sets of variables $\mathbf{u}$,
$\mathbf{v}$, we define
\begin{align}
F(\mathbf{u},\mathbf{v})\equiv\prod_{u_i\in\mathbf{u},
\;v_j\in\mathbf{v}}F(u_i,v_j),\quad
F^>(\mathbf{u},\mathbf{v})\equiv\prod_{i>j\atop
u_i\in\mathbf{u},\;v_j\in\mathbf{v}}F(u_i,v_j)\;.
\end{align}

Proportionality factor between ABA and CBA Bethe state:
$|\uu\rangle_{\text{alg}}= c_\uu \, |\uu\rangle_{\text{cor}}$ is given
by
\be c_\uu&=& i^{N+M} \ \prod_{a=1,2}\prod_{j<k} {u_{a,j}-u_{a,k}
+i\over {u_{a,j}-u_{a,k} }}.  \ee \label{cu}
While the formula (\ref{c123}) can be explicitly used for a small
numbers of magnons, it is not adapted for taking the classical limit
where the number of magnons is large.  The main obstruction for taking
the classical limit of (\ref{c123}) is that the scalar products
involved are between off-shell states, and there is no closed form
expression such as a determinant for this scalar product.  In the
following sections, we restrict our attention to a particular
situation where the 3-point function can be written in terms of a
scalar products of an off-shell state and an on-shell state.

\section{The $su(2)$ cubic vertex in terms of scalar products }

\la{sect:3pf=scpr} 
\la{sect:su2.3pf=scpr}

In preparation for the computation of the type-$\{2,3,3\}$ $su(3)$
3-point function that we are interested in, we review an analogous
computation of a type-$\{2, 2, 2\}$ $su(2)$ 3-point function in
\cite{Omar}.  Consider the 3-point function of the operators $\CO_i$,
of lengths $L_i$, and rapidities $\uu_i$ with cardinalities $N_i$, $i
\in \{1, 2, 3\}$.  In the following we set $\uu_1 = \uu$, $\uu_2 =
\vv$, and $\uu_3 = \ww$.  In our conventions, $\CO_1$ consists of the
fundamental fields $\{ Z, X\}$, $\CO_2$ of $\{\bar Z,\bar X\}$, and
$\CO_3$ of $\{Z, \bar X\}$.

It is advantageous to generalize the problem slightly by introducing
inhomogeneities associated with the sites of the three spin chains.
Thus the $i$-th chain is characterized by inhomogeneities $
\thth^{(i)}=\{ \th^{(i)}_1, \dots, \th^{(i)}_{L_i}\}$, $i=1,2,3$.  The
three sets of inhomogeneities are not independent, because the
inhomogeneities associated with two sub-chains whose fundamental
fields are contracted should match.  The independent inhomogeneities
associated with the contractions between the $i$-th left sub-chain and
the $j$-th right sub-chain are denoted by $\thth^{(ij)}$.  The
cardinality of the set $\thth^{(ij)}$ is $L_{ij}$.  In this notation
 \be \thth^{(1)} = \thth^{(12)}\cup \thth^{(13)}, \ \thth^{(2)} =
 \thth^{(12)}\cup \thth^{(23)}, \ \thth^{(3)} = \thth^{(13)}\cup
 \thth^{(23)} .  \ee

The planarity of the $\langle Z \bar Z\rangle$ contractions between
the operators $\CO_2$ and $\CO_3$ and the $\langle X\bar X\rangle$
contractions between the operators $\CO_1$ and $\CO_3$ selects the
component of $\CO_3$ with $N_3= L_{13}$ successive $\bar X$'s and
$L_3-N_3$ successive $Z$'s, as in Fig.  \ref{fig:3pf-SU3}.
Consequently, the correlation function is given by the product of two
factors:
 \begin{itemize}
 \item
 The probability to find the component 
 $\Tr [ Z^{L_{23}} \bar X^{L_{13}}]$   in the state $|\ww\rangle$.
 \item
The contribution of the remaining contractions can be recast as the
scalar product of an on-shell state of rapidities $\uu$ and an
off-shell state of rapidities $\vv$, in a spin chain of length $L_1$.
 \end{itemize}
We present below the derivation of the two factors using the language
of the six-vertex model.

\subsection{The $su(2)$ Bethe states as six-vertex-model partition
functions} 
\begin{figure}[h]
         \centering
                 \begin{minipage}[t]{0.55\linewidth}
            \centering
             \includegraphics[scale=0.6]{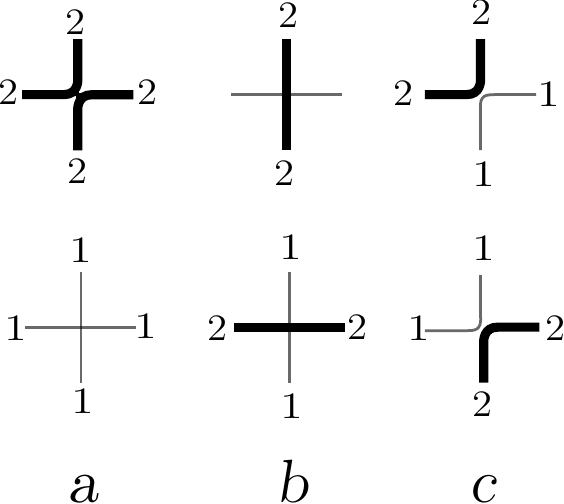}
	  \caption{\small \small Graphical representation of the six
	  non-zero elements of the $L$ -matrix \re{Rmatrix}.  The
	  rapidity $u$ is associated with the horizontal line, while
	  the rapidity $z= \th +i/2$ is associated with the vertical
	  line.}
\label{fig:6vertex}
         \end{minipage}
           \end{figure}

The three type of vertex configurations, $a$, $b$, $c$
represented in Fig.  \ref{fig:6vertex} have weights
 \be \la{abc} a(u-z)=\frac{u-z+i}{u-z}\;, \quad b(u-z)=1\;, \quad
 c(u-z)=\frac{i}{u-z}.  \ee
The rapidities $u$ and $z= \th + i/2$ are associated respectively with
the horizontal and with the vertical lines.  These weights are given
by the three types of non-zero elements of the $L$-matrix, which in
our case coincides with the $R$-matrix, \be \la{Rmatrix} {L}(u-z)=
{\bf I} + \frac{i }{u-z}\ {\bf P} \; ,\qquad z= \th + i/2.  \ee
 Here ${\bf I}$ is the identity matrix and ${\bf P}$ is the
 permutation matrix.

% \vskip 0.8cm

%\vskip 0.8cm

Consider the expansion of a Bethe vector $|\uu\rangle$ in the local
basis $|s_1, \dots, s_L\rangle$, where $s_k \in \{1, 2\}$,
 \be \la{Bethestate} |\uu\rangle =\sum_{s_1,\ldots,s_L=1,2}
 \psi_{s_1,\ldots,s_L}(\uu)|{s_1,\ldots,s_L}\rangle\, .  \ee
 Each of the components $\psi_{s_1,\ldots,s_L}(\uu)$ is a sum over all
 the possible vertex configurations with on a $L\times N$ rectangle,
 with all indices fixed to 1 on the left and the upper boundaries, 2
 on the right boundary, and free indices equal to $s_1,\ldots,s_L$ on
 the lower boundary, as shown in Fig.  \ref{fig:su2Bethestate}.
 Similarly the dual Bethe state is represented by the partition
 function of the six-vertex model on a rectangle, with boundary
 conditions 1 on the right and the lower boundary, and 2 on the left
 boundary.
 
 \subsection{The $su(2)$ scalar/ inner product in terms of the
 6-vertex model}

\begin{figure}[H] 
 %    \vskip 1cm
         %%----start of first figure----
	 \begin{minipage}[t]{0.50\linewidth}
            \centering
            \includegraphics[width=5.9 cm]{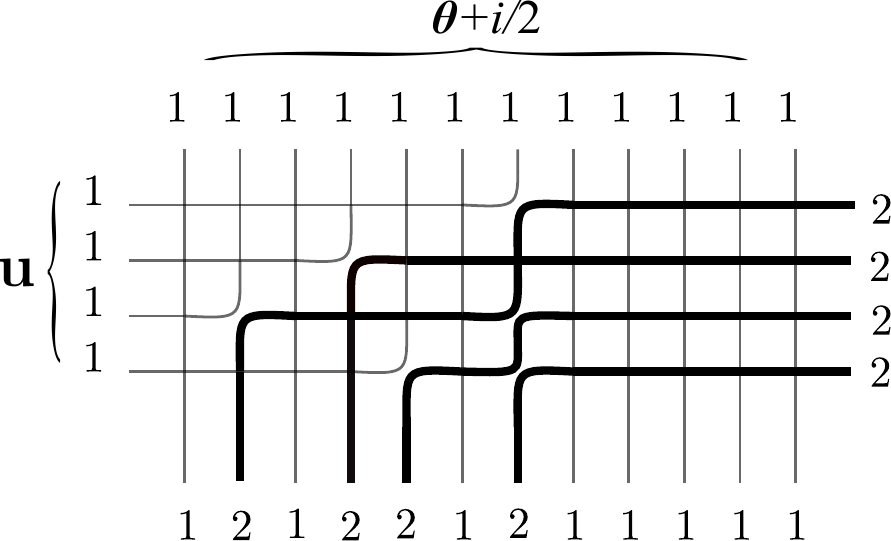}
	   \caption{ \small A six-vertex configurations for the
	   coefficient $\psi_{_{121221211111}}(\uu)$ of the Bethe
	   state $| \uu\rangle$, Eq.  \re{Bethestate}.  }
  \label{fig:su2Bethestate}
         \end{minipage}%
         \hspace{1.0cm}%
                \begin{minipage}[t]{0.41\linewidth}
            \centering
            \includegraphics[width=5.9cm]{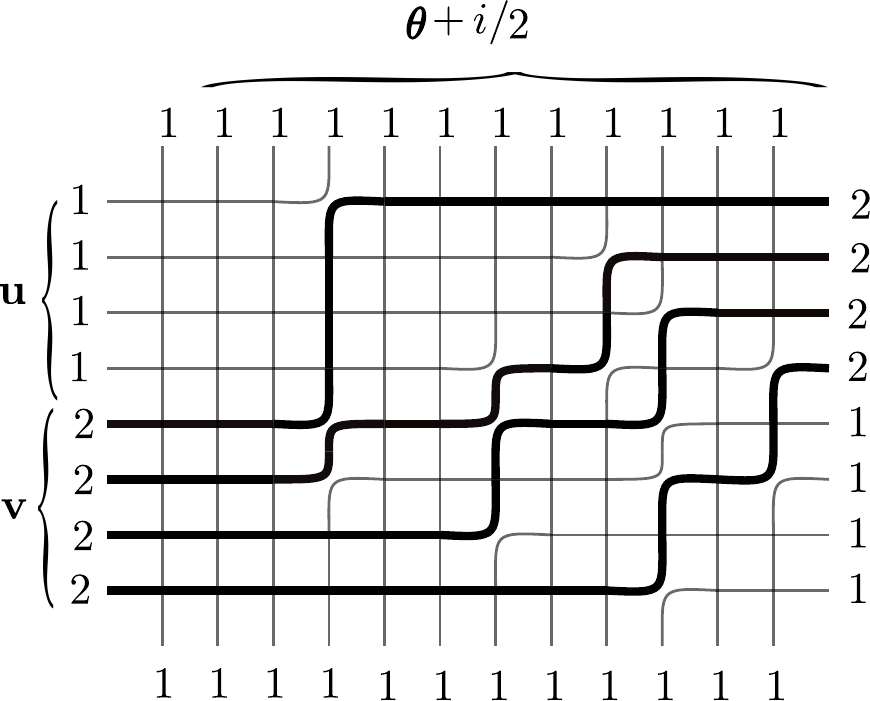}
	     \caption{ \small A six-vertex configurations for the
	     inner product $\<\vv, \uu\>$.  }
\label{fig:6vscprod}
         \end{minipage}
      \end{figure}

 With the normalisation of the basis
  \be
  \< s_1, \dots, s_L|r_1,\dots, r_L\>=\d_{s_1,r_1}\dots   \d_{s_L,r_L}\, ,
  \ee
  the scalar product of two (in general off-shell) Bethe states
  \be \la{scalprsu2} \langle \vv|\uu\rangle = \sum_{s_1,\dots, s_L
  =1,2} \psi _{s_1\dots s_L} (\uu) \ \psi_{s_1\dots s_L}(\vv)^* \ee
 is obtained simply by gluing two such partition functions, as shown
 in Fig.  \ref{fig:6vscprod}, and summing over the free indices.  In
 addition to the sesquilinear form \re{scalprsu2}, which is {\it \lq
 the scalar product\rq\ }, we define a bilinear form, which we call
 {\it \lq the inner product\rq\ }
  \be \la{bilinearsu2} \langle \vv,\uu\rangle= \langle \vv, \uu\rangle
  = \sum_{s_1,\dots, s_L =1,2} \psi_{s_1\dots s_L} (\uu) \
  \psi_{s_L\dots s_1}(\vv).  \ee
  The vertex representation of the inner product \re{bilinearsu2} is
  obtained by gluing two lattices as the one shown in Fig.
  \ref{fig:su2Bethestate} and summing over the free spin indices.  The
  result is the six-vertex partition function on a lattice with
  indices $1^L$ on the upper and lower boundaries, and $1^{N}2^N$ on
  the left and right boundaries in Fig.  \ref{fig:6vscprod}.  The
  symmetry $\langle \vv,\uu\rangle= \langle \vv, \uu\rangle $ of the
  inner product follows from the symmetry of the weights $a,b,c$ of
  the vertices in Fig \ref{fig:6vertex} with respect to a rotation by
  180 degrees.
  
  \vskip 0.6cm

%%%%%%%%%%%%%%%%%%%%%%%%%%%%%
\vskip 1.0cm

It follows from the hermitian conjugation properties of the creation
($B$) and the annihilation ($C$) operators (see the historical note
\cite{2009arXiv0911.1881K}) that for $N$-magnon states
  \be
  \langle\vv|\uu\rangle = 
  (-1)^N \ \< \vv^*, \uu\>,
  \ee
where the set of rapidities $\vv^*$ is obtained from $\vv$ by complex
conjugation.  Since the Hamiltonian of the XXX chain is hermitian, the
sets of rapidities of the Bethe eigenstates are symmetric under
complex conjugation.  Therefore the normalisation factor $\CN_1 =|
\<\uu, \uu\>| $ in \re{twopf} is equal to the (squared) norm $\< \uu
|\uu\>$ of the Bethe eigenstate.

The structure constant $C^{(0)}_{123}$ is equal, up to the
normalization factor, to the cubic vertex made of the wave functions
of the Bethe states in the representation \re{Bethestate},
\be \la{C123cubic} C^{(0)}_{123} ={ \< \uu, \vv, \ww\>\over \sqrt{\<
\uu,\uu\>\< \vv, \vv\>\<\ww,\ww\>}} \, , \ee
where the form of the cubic vertex depends on the choice of the three
$su(2)$ sectors.  In our particular case
 \be 
  \< \uu, \vv, \ww\>&\equiv& \sum _{ } \psi^{ }_{{
 \underbrace{_{1\dots 1}}_{L_{23} }} s_{{L_{12}}} \dots s_1 } (\vv)\
 \psi^{ }_{s_1 \dots s_{L_{12}} \underbrace{_{2\dots 2}}_{L_{13} }
 }(\uu)\ \ \psi _{ \underbrace{_{2\dots 2}}_{L_{13} }
 \underbrace{_{1\dots 1}}_{L_{23} }}(\ww) \, , \la{cubicvx}
   \ee  %
  where the summation indices $s_1, \dots s_{L_{12}} $ take values $1$
  and $2$.

\vskip 0.3cm

\begin{figure}[H]
         \centering
                 \begin{minipage}[t]{0.80\linewidth}
            \centering
         \includegraphics[scale=0.63]{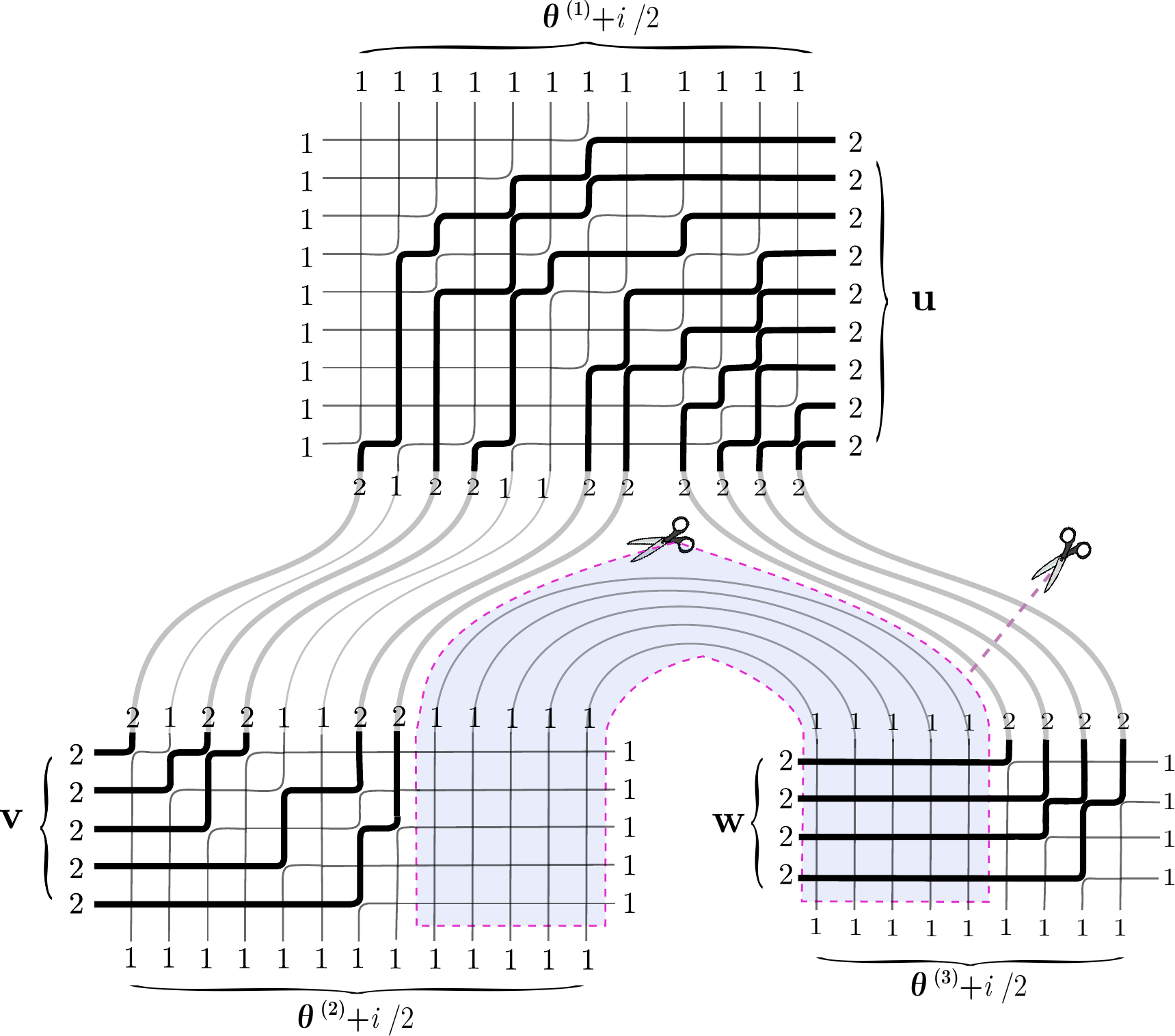}
\caption{\small \small The $su(2)$ cubic vertex in terms of six-vertex
configurations.  The shaded area factorizes out and can be cut out.
Furthermore, the right piece is connected to the rest only with type-2
contractions and factorizes out.  The lattice splits into two pieces
which do not talk to each other and can be evaluated separately.  }
 \label{fig:TAILOR}
 \vspace{1.0cm}
         \end{minipage}
          \end{figure}

 The cubic vertex $\< \uu, \vv, \ww\>$ can be evaluated using the fact
 that it gives the partition function of the six-vertex model on a
 lattice obtained by gluing three rectangular lattices with dimensions
 $L_1\times N_1$, $L_2\times N_2$ and $L_3\times N_3$ as shown in Fig.
 \ref{fig:TAILOR}.  The indices $1$ and $2$ are identified with $Z$
 and $X$ or their complex conjugates, depending on the operator under
 consideration.  First we notice that in the part of the lattice that
 has vertical lines labeled by $\thth^{(23)}$, represented by the
 shaded area in Fig.  \ref{fig:TAILOR}, there is only one six-vertex
 configuration, and therefore its contribution to the cubic vertex
 factorizes out.  The factor is a pure phase if the sets $\vv$ and
 $\thth^{(23)}$ are symmetric under complex conjugation.  We will
 assume that this is the case and will ignore this phase factor.
 Therefore we can delete this part of the lattice.

\begin{figure}[H]
  \centering
                         \begin{minipage}[t]{0.80\linewidth}
            \includegraphics[width=10.9cm]{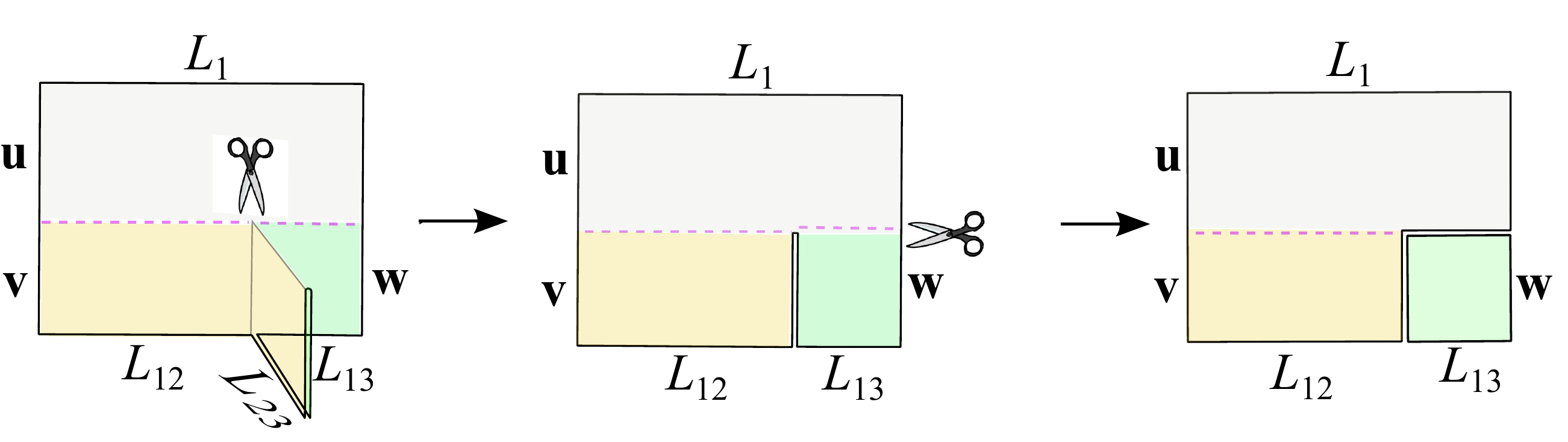}
\caption{ \small Schematic representation of the lattice obtained by
gluing the rectangular lattices corresponding to the states $\langle
\uu|$, $|\vv\rangle$ and $| \ww\rangle$, with subsequent removal of
the redundant piece and separating the two non-interacting
sub-lattices.  After the removal of the redundant piece, the states
$|\vv\rangle$ and $|\ww\rangle$ are no more Bethe eigenstates, because
the chain are shortened ($L_1\to L_{12}$ and $L_3\to L_{23}$).}
\label{fig:cutting6v}
         \end{minipage}
      \end{figure}

 Next, we observe that the sub-lattice associated with the operator
 $\CO_3$ factorizes because all lines that connect it with the rest of
 the lattice are of type 2.  (This factorisation is obvious in the
 expression \re{cubicvx} for the cubic vertex.)  These operations are
 schematically represented in Fig.  \ref{fig:cutting6v}.

The problem boils down to the calculation of two independent
six-vertex partition functions, which give the two non-trivial factors
in the structure constant.  These two factors will be computed using
the {\it freezing procedure}.  The freezing procedure for the first
factor works as follows.  One starts from a rectangular lattice
corresponding to the scalar product $\< \tilde \vv|\uu\>$.  Both sets
of rapidities have cardinality $N_1$.  The first $N_2$ rapidities
$\tilde \vv$ coincide with the rapidities $\vv$ characterizing the
operator $\CO_3$, the rest $N_3=N_1-N_2 =L_{1} - L_{12}$ of the
rapidities $\tilde \vv$ will be denoted by $ \tilde v_{N_2+1} = \tilde
z_{L_{12}+1}, \dots, \tilde v_{ N_1} = \tilde z_{L_{1}}$, or
symbolically, $\tilde \vv = \vv\cup\tilde \zz$.

 \begin{figure}[h!]
                \begin{minipage}[t]{0.89\linewidth}
            \centering
            \includegraphics[width=12.9cm]{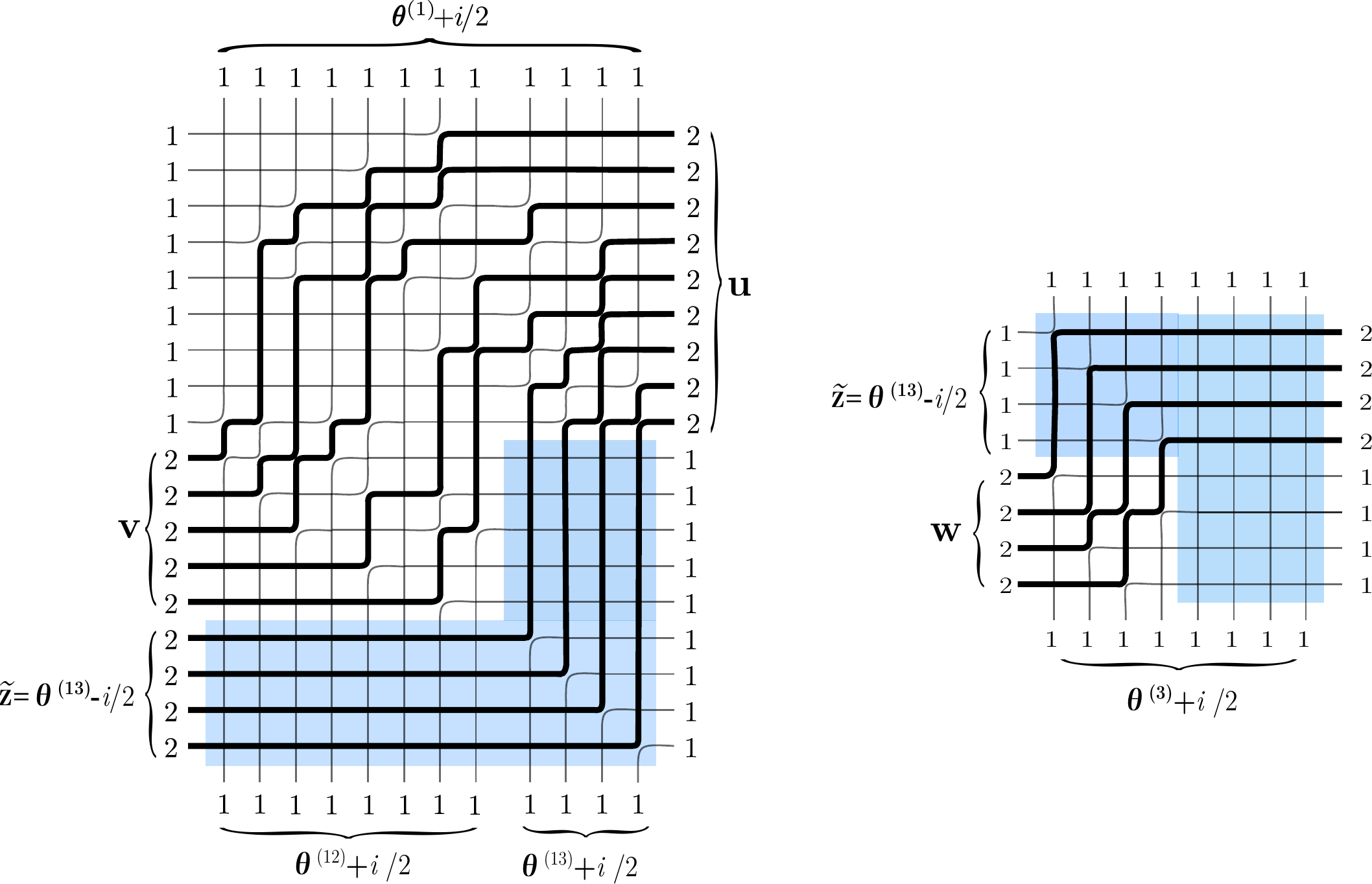}
	     \caption{ \small The freezing procedure for the two
	     factors in $C_{123}^0$ }
\label{fig:noodsw-Det}
         \end{minipage}
      \end{figure}

 \subsection{The $su(2)$ freezing procedure}
If we adjust the rapidity of the last magnon to the value of the last
inhomogeneity, $ \tilde z_{L_{1}} = \th_{L_1} - i/2$, then the vertex
at the low right corner is necessarily of type $c$.  Then the only
possibility for the rest of the vertices on the last row and the last
column is that they are type $b$.  This is what we call ``freezing".
Hence last row and the last column form a hooked index line carrying
the index $2$, as shown in Fig.  \ref{fig:noodsw-Det}, left.  This
procedure is repeated $N_3$ times, the rapidities of the lowest $N_3$
rows fixed to $\tilde \zz = \thth^{(13)} - i/2$.  The result is that
the rightmost $N_3$ indices below the lowest horizontal $u$-line are
fixed to the value $2$.  After removing the frozen part of the
lattice, shaded in blue in the figure, we obtain that the first factor
in the cubic vertex equals the scalar product $\< \vv\cup \tilde \zz,
\uu\>$.  The contribution of the frozen part of the lattice is the
product of all $c$-vertices on the diagonal, which equals
$(-1)^{L_{13}}$, a factor we will ignore.

In a similar way we compute the second factor .  The freezing
procedure is shown in Fig.  \ref{fig:noodsw-Det}, right.  We start
with a scalar product $\< \ww, \tilde \zz\>$ for a chain of length
$L_3= N_3 + L_{23}$.  We freeze the rapidities of the bra state to
$\tilde \zz = \thth^{(13)} - i/2$.  The frozen area (shaded in blue)
gives a contribution, which is a pure phase if both sets $\ww$ and
$\thth^{(23)}$ are symmetric under complex conjugation.  We will
assume that this is the case and will ignore this factor.  The rest of
the lattice gives the second factor in the expression for the cubic
vertex.  We find
  \be \< \uu, \vv, \ww\>= \langle\vv \cup \tilde \zz,
  \uu\rangle_{\thth^{(1)}} \ \langle \ww, \tilde \zz
  \rangle_{\thth^{(3)}} \ee
up to a factor which takes into account the contribution of the
deleted and added pieces of the lattice.  This factor is a pure phase,
since the set $\ww$ is symmetric under complex conjugation, and can be
ignored.

   \section{The $su(3)$ cubic vertex   in terms of scalar products } 
 
\la{sect:su3.3pf=scpr}

 As before, we consider that the three operators, $\CO_1,\ \CO_2,\
 \CO_3$, are described by three sets of rapidities
 $\uu=\{\uu_1,\uu_2\},\ \vv=\{\vv_1,\vv_2\} $ and $
 \ww=\{\ww_1,\ww_2\}$ with cardinalities respectively $N_1+ M_1, N_2+
 M_2 $ and $ N_3+M_3$.  In the configuration we are considering,
 $\ww_2=\emptyset$, since $\CO_3$ is an $su(2)$ operator.  We refer to
 Section \ref{sect:dilog.su3} for the equations obeyed by these
 rapidities.

 \smallskip
 
 Again, we have two types of contributions to the correlation
 function:

 \begin{itemize}
 \item
 the contribution of the $\langle Z \bar Z\rangle$ contractions
 between the operators $\CO_2$ and $\CO_3$, through the factor
 $\langle\tilde \zz_1,\ww_1\rangle$, with $\tilde
 \zz_1=\thth^{(13)}-i/2$ and $\# \tilde \zz_1 = N_3$,
 \item
the remaining contractions, which can be recast as the inner product
$\langle\vv \cup\tilde \zz , \uu\rangle$ between an on-shell vector of
a spin chain with length $L_1$ and rapidities $\uu=\{\uu_1,\uu_2\}$
and an off-shell state with the same length and rapidities
$\(\vv=\{\vv_1,\vv_2\} \)\cup\( \tilde \zz=\{\tilde \zz_1,\tilde
\zz_2\}\)$, with $\tilde \zz_1= \thth ^{(13)} -i/2 $ and $\tilde
\zz_2=\thth ^{(13)} -i$.
 \end{itemize}

Below we evaluate, using the freezing argument, the $ \{2,3,3\}$ type
structure constant (Fig.  \ref{fig:3pf-SU3}),
\be \la{C123cubicsu3} C^{(0)}_{123} ={ \< \uu, \vv, \ww_1\>^{\su3}\over
\sqrt{\< \uu, \uu\>^\su3\< \vv, \vv\>^\su3\<\ww_1, \ww_1\>}} \, .  \ee
We will show that the corresponding cubic vertex is  given by
  \be \la{cubicsu3} \< \uu, \vv, \ww\>^{\su3}&=& \langle\vv \cup\tilde
  \zz, \uu\rangle_{\thth^{(1)}}^{^\su3 } \ \langle \ww_1, \tilde \zz_1
  \rangle_{\thth^{(3)}} \\
 \no  \\
 {\rm with}\quad \quad\tilde \zz & =& \{ \tilde \zz_1,\tilde \zz_2\}=
 \{ \thth ^{(13)} -i/2 ,\ \thth ^{(13)} -i\} .  \ee
Here $ \langle \ \ , \ \ \rangle$ denotes, as before, the $su(2)$
inner product, and $\langle \ \ , \ \ \rangle^\su3$ denotes the
$su(3)$ inner product.

\subsection{The $su(3)$ Bethe states in terms of the  15-vertex-model}
\begin{figure}[h!]
         \centering
                 \begin{minipage}[t]{0.7\linewidth}
            \centering
             \includegraphics[scale=0.60]{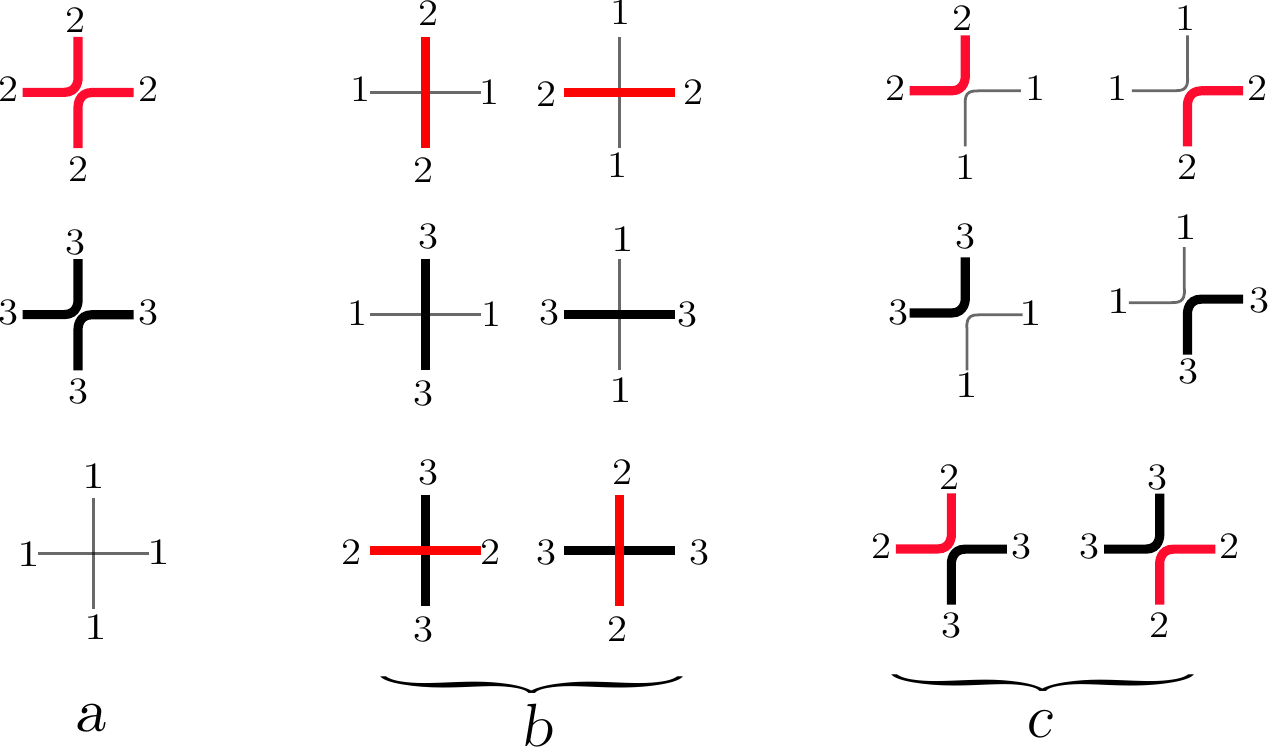}
	  \caption{ \small Graphical representation of the 15
	  non-zero elements of the $su(3)$ $L$ -matrix, Eq.
	  \re{Rmatrix}.  The rapidities $u$ and $z= \th +i/2$ are
	  associated with the horizontal and the vertical lines,
	  respectively.}
\label{fig:6vertex15}
         \end{minipage}
           \end{figure}

 In order to generalize the freezing procedure to
 $su(3)$, let us first show how to represent the components of the
 $su(3)$ Bethe vectors in terms of configurations of a 15-vertex model
 shown in Fig.  \ref{fig:6vertex15}.  The vertices are similar to
 those from Fig.  \ref{fig:6vertex}, with the difference that the
 indices carried by the lines can be now $1, 2$ or $3$.  We represent
 them graphically by thin, red and black lines, respectively.  The
 weights are identical to those from equation (\ref{abc}), depending
 on whether the indices carried by the lines are equal or different.
 
 \bigskip

 \bigskip

 \noindent
 The Bethe vector $|\uu\rangle$ is given by the expansion \be
 \la{su3bethev} |\uu\rangle =\sum_{s_1,\ldots,s_L=1}^3
 \psi_{s_1,\ldots,s_L}(\uu) \ |{s_1,\ldots,s_L}\rangle \ee where
 $\psi_{s_1,\ldots,s_L}(\uu)$ is a sum over all the possible 15-vertex
 configurations on a rectangular lattice with $L_1+N_1$ vertical lines
 and $N_1+M_1$ horizontal lines, with the free spin indices equal to
 $s_1,\ldots,s_L$.  An example for such a vertex configuration is
 given in Fig.  \ref{fig:su3noodle}.  The first $L_1$ vertical lines
 carry rapidities $\theta^{(1)}_1+i/2,\ldots,\theta^{(1)}_{L_1}+i/2$
 and spin indices $1$ on the top, which correspond to the vacuum
 $|\Omega\rangle= |1^L\rangle \equiv|11\ldots 1\rangle $.  The right
 $N_1$ vertical lines carry rapidities $u_{1,1}+i\ldots u_{1,N_1}+i$
 and have index $2$ on the top.  At the bottom, the first $L_1$
 indices are free, and the last $N_1$ ones are fixed to $1$.  The
 lower $N_1$ horizontal line correspond to the first-level magnons and
 carry rapidities $u_{1, 1}, \dots, u_{1, N_1}$.  The higher $M_1$
 horizontal lines represent the second-level magnons with rapidities
 $u_{2,1},\dots, u_{2, M_1}$.  Due to the particular spin and rapidity
 choices, the shaded regions are frozen to the particular
 configuration shown in the Figure.  This diagram is equivalent to
 (a special case of\footnote{In  \cite{reshet-SU3},  there are
  had two momentum-carrying
nodes,  while our spin chain has only one momentum-carrying node.}) the
 one used by Reshetikhin in \cite{reshet-SU3}.

\begin{figure}[h!]
         \centering
                 \begin{minipage}[t]{0.8\linewidth}
            \centering
       \includegraphics[scale=0.6]{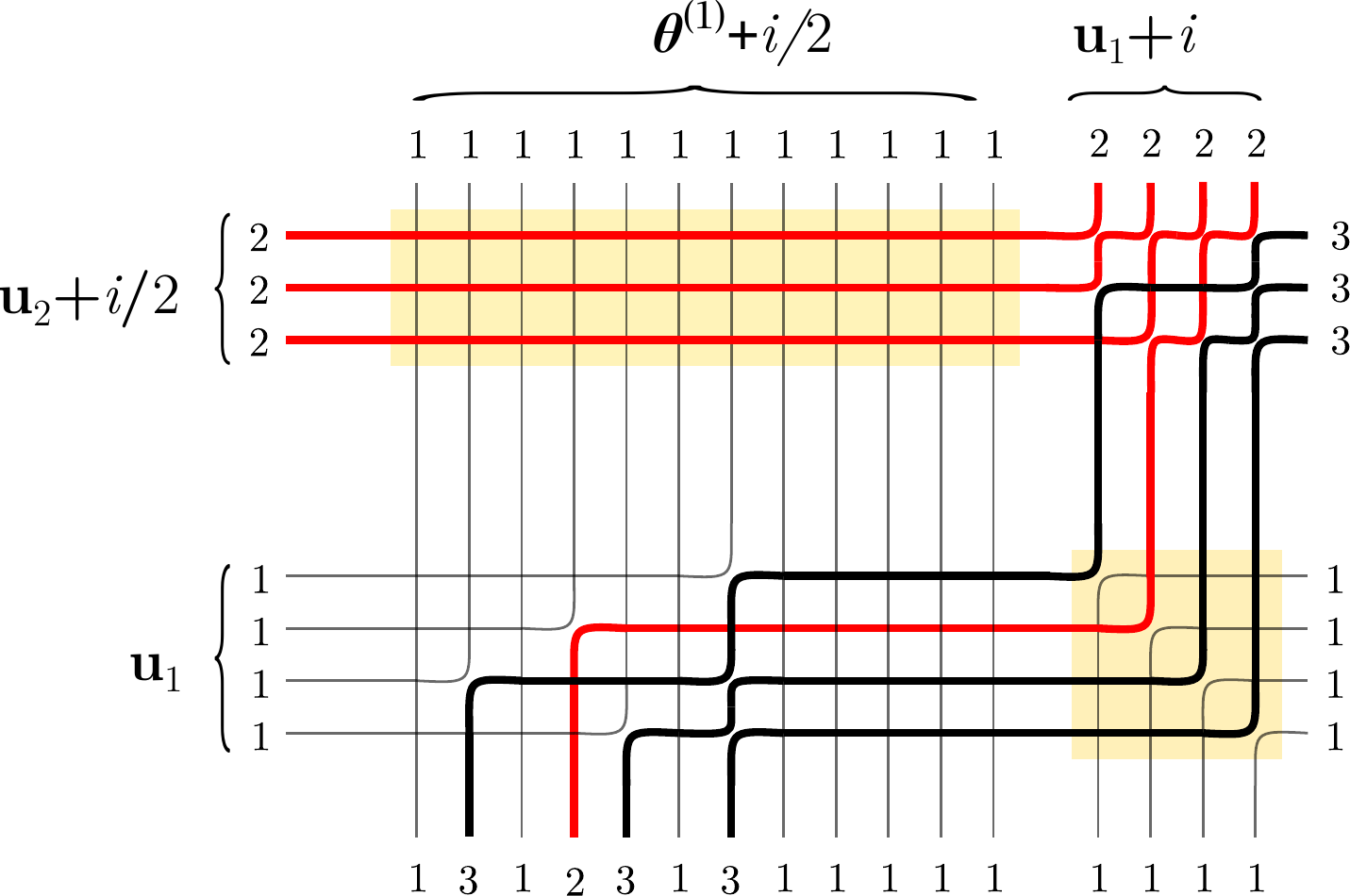}
	 \caption{\small A configuration contributing to the
	 coefficient $\psi_{131231311111}(\uu)$
   %in front of $|131231311111\rangle$
 of a $su(3)$ Bethe vector \re{su3bethev} with $L=12$, $N=4$ and $M=3$
 .} \la{fig:su3noodle}
         \end{minipage}
           \end{figure}

 \begin{figure}[h!]
 \begin{center}
 \includegraphics[scale=0.7]{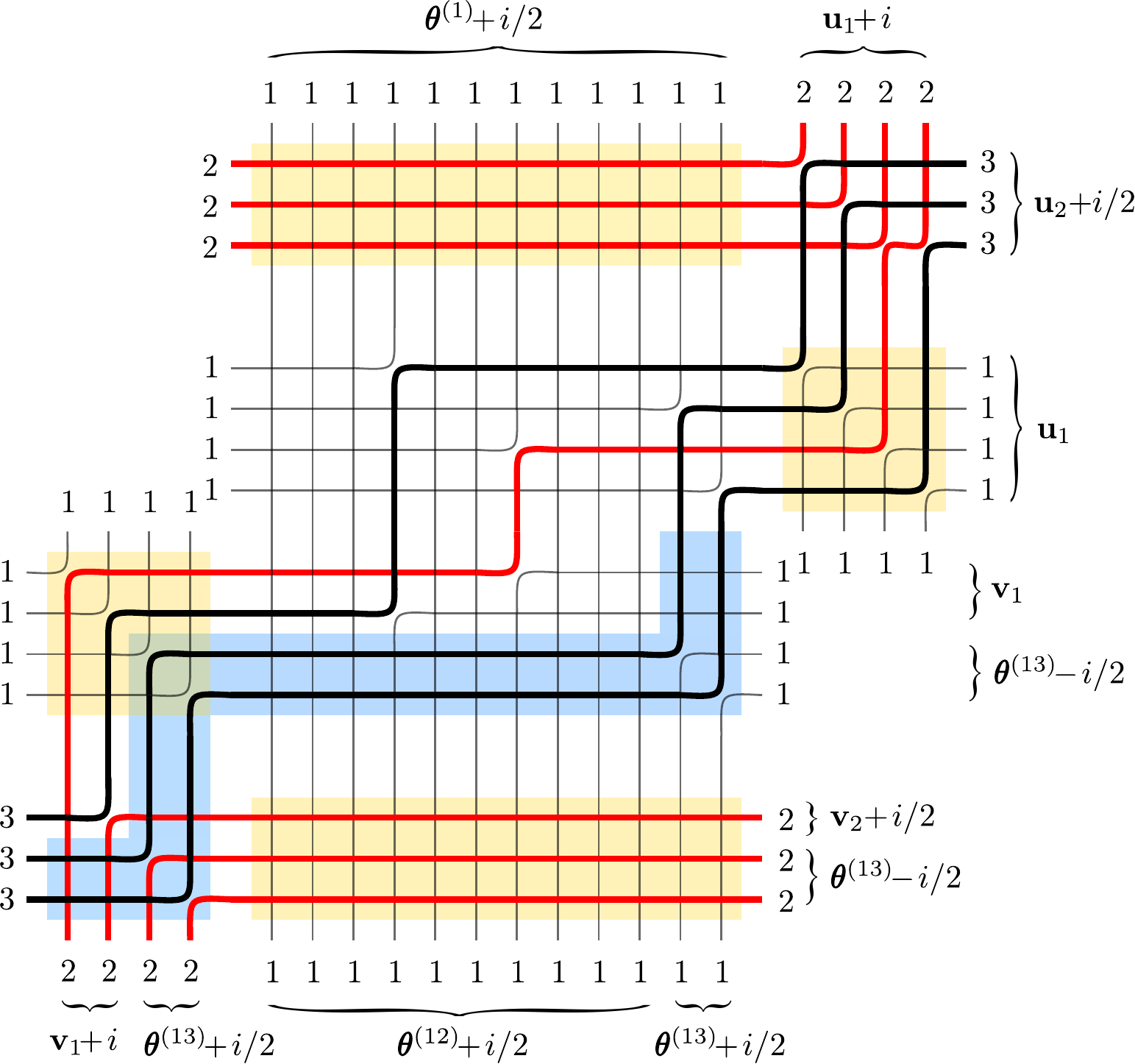}
 \caption{\small The inner product $\langle \tilde \vv, \uu\rangle$
 and the freezing to $\langle \vv\cup \thth^{(13)}- i/2, \uu\rangle$
 .}
 \label{fig:su3ScalarProduct}
 \end{center}
 \end{figure}
 
\bigskip
 
The structure constant factorizes as in the $su(2)$ case.  The two
factors can be cast in the form of scalar products of an on-shell and
an off-shell Bethe states by applying the freezing procedure.

\bigskip

\subsection{The $su(3)$ freezing procedure}

 Consider the scalar
product of two $su(3)$ states of the first chain, $\langle \tilde \vv,
\uu\rangle$=$\langle \{\tilde \vv_1,\tilde \vv_2\},
\{\uu_1,\uu_2\}\rangle$, as represented in figure
\ref{fig:su3ScalarProduct}.  Our purpose is to freeze the rightmost
$N_3$ indices to the value $3$, as imposed by the planarity of the
contractions in the three point function.  This can be done by setting
the last $N_3$ rapidities of first level magnons at their freezing
values,
 \be
 \tilde v_{1,N_2+1}= \th^{(13)}_1- i/2, \  \ldots, \  \tilde v_{1,N_1}=
 \th^{(13)}_{N_3} -i/2.
 \ee
 This will insure that the corresponding frozen region contains only
 red and black lines propagating from the top to the bottom of the
 diagram.  The number of black or red lines is not fixed by the
 freezing, only their sum is fixed.  In order to force all the lines
 in the frozen region to be black, we have to apply once again the
 freezing procedure to the magnons of second level, by fixing
 \be
 \tilde v_{2,M_2+1}=  
% z_{2,1} 
% \equiv 
 \th^{(13)}_{1} - i, \ \ldots \ \tilde v_{2,M_1}= \th^{(13)}_{N_3} -i.
 \ee
  The remaining magnons are set to the corresponding values in the
  state $|\vv\rangle$,
 \be
 \tilde v_{1,1}=v_{1,1},\ \ldots,\ \tilde v_{2,M_2}=v_{2,M_2}
.
\ee
 This gives us the first factor of the expression \re{cubicsu3} for
 the cubic vertex.  The second factor, $\< \zz_1, \ww_1\>$, is the
 same as in the $su(2)$ case.

\section{The $su(2)$ structure constant in terms 
of the $\caA$-functional} 

\la{sect:dilog.su2}

In the $su(2)$ case, the rapidities of the on-shell Bethe states
satisfy the Bethe equations
\be \la{BAEXXX}
  \prod_{l=1}^L  {u_j - \th_l+ i/2 \over u_j-\th_l - i/2}=-
 \prod_{k =1}^N {u_j - u_k +i\over u_j -u_k - i} \, , \quad j=1, \dots, N .
 \ee
 The Bethe equations \re{BAEXXX} follow from the requirement that the
 eigenvalue $T(u)$ of the $T$-matrix,
\be
\la{Tmatrix}
T(u)={Q_\thth^+   (u)\over
Q^-_\thth(u)} {Q^{--}_\uu(u)\over Q_\uu(u)}
+
 {Q^{++}_\uu(u)\over Q_\uu(u)} 
\equiv  e^{ip_\uu^1(u)} + e^{ip^2_\uu(u)}
\ee
has vanishing residues at the Bethe roots.\footnote{With the
normalisation \re{Rmatrix} of the $L$-matrix, $T(u)$ is not a
polynomial, but has poles at $u-\th_l$.} Here $Q_\uu$ and $Q_\thth$
are the Baxter polynomials
\be \la{defQ} Q_\uu(u) = \prod_{j =1}^N (u- u_j ) , \qquad
Q_{\thth}(u)= \prod_{l =1}^L (u- \th_l), \ee
and  
\be\la{shifts} Q^\pm (u)= Q(u\pm i/2), \ \ Q^{\pm\pm} (u) = Q(u\pm i),
\quad Q^{[n]}(u) = Q(u+i n/2).  \ee

The Bethe equations read
 \be \la{BAE} 
 e^{2ip_\uu(u_j)}=-1, \qquad  j=1,\dots, N, \ee
 where the pseudomomentum $p_\uu $, known also as the counting
 function, is defined modulo $\pi$ by
\be \la{defquazimomentum} e^{2i p_\uu } = e^{ i p^1_\uu-i p^2_\uu }=
{Q_\thth^- \over Q^+_\thth }\ {Q^{++}_\uu\over Q^{--}_\uu} \, .  \ee

The $T$-matrix \re{Tmatrix} is normalized according to the vertex
representation with weights \re{abc}.  The eigenvalues of the diagonal
elements of the monodromy matrix on the vacuum, $A(u)$ and $D(u)$, are
given by
  \be
  \begin{aligned}
  D(u) &= e^{ip^1_{\emptyset} (u)}=\prod_{l=1}^L b(u - \th_l - i/2) =1
  , \\
  A(u) &= e^{ip^2_{\emptyset} (u)} =\prod_{l=1}^L a(u - \th_l - i/2) =
  { Q_\thth ^+( u)\over Q^-_\thth ( u)} .
 \end{aligned}
  \ee

 The inner product of an on-shell Bethe state $\uu$ and an off-shell
 state $\vv$ is given in a determinant form by
 \cite{slavnov-innerpr}\cite{nikitaslavnov}
   \be \la{defSuv0}
   \< \vv,\uu\> %=  \< \uu,\vv\> 
  &=&\ \prod_{j =1}^N A_\thth(v_j) \ \caS_{\uu, \vv}\, , \la{defscpr}
  \ee where
 \be \caS_{\uu, \vv} &=& {\det_{j k}\O(u_j , v_k ) \over \det_{j k}
 {1\over u_j -v_k +i} } \la{defSuv}\, , \ee is the Slavnov
 determinant.  The Slavnov kernel $\O(u,v)$ is defined by\footnote{In
 order to simplify the formulas, here (as well as in
 \cite{3pf-prl}\cite{SL}\cite{sz}) we use a different normalisation
 for the Slavnov matrix than in \cite{nikitaslavnov}.  In our
 conventions, the Slavnov kernel depends only on the pseudomomentum
 $p_\uu=p_\uu^2-p_\uu^1$.}

  \be \la{defOhat} \O(u,v)= t(u-v) - e^{2 i p_\uu(v)} \ t(v-u)\, ,
  \qquad t(u) = {1\over u } - {1\over u+i}\, .  \ee

The Slavnov determinant cannot be directly evaluated in the classical
limit.  This can be done using the representation in terms of the
$\caA$-functional introduced in \cite{GSV}.  The Slavnov determinant
$\caS_{\uu,\vv}$ was expressed in terms of the $\caA$-functional first
for the limit $\uu\to\infty$ \cite{GSV} and then in the general case
\cite{3pf-prl}\cite{SL}.  Later a more compact expression was found in
\cite{sz}.

The $\caA$-functional, whose properties are listed in Appendix
\ref{appendix:caA}, is defined for any function $f(u)$ and any set of
points in the complex plane $\uu= \{ u_1, \dots, u_N\}$ as follows,
 \be
 \la{detformula}
 \caA^\pm _\uu[f] &=& { \det_{j k} \( u_j ^{k-1} - f(u_j ) \, (u_j \pm
 i)^{k-1}\) \over \det_{j k} \( u_j ^{k-1}\) } \, . 
 \ee
The Slavnov determinant is expressed in terms of this functional as
\cite{sz} \footnote{The expression \re{tilSLA} for the Slavnov
determinant depends on the ensemble of the rapidities $\uu$ and $\vv$
in a completely symmetric way.  This remarkable symmetry of this
expression follows from the fact that, due to the global $su(2)$
symmetry, the annihilation operators with rapidities $\vv$ can be
replaced by creation operators with the same rapidities and a global
raising operator \cite{sz}.  See also the exercises of the 3rd day of
the 4th Mathematica School (http://msstp.org/?q=node/272).  }
  \be \la{tilSLA} \caS_{\uu, \vv} = (-1)^N\ \caA^+_{\uu\cup\vv}[
  {Q^-_\thth \over Q^+_\thth} ] .  \ee

Let us express the scalar products in the expression for the 
 cubic vertex  in terms of the $\caA$-functional.  
 We find for the two inner products  in  \re{cubicsu3}
 \be \la{firstfactor} \langle\vv \cup \tilde \zz,
 \uu\rangle_{\thth^{(1)}} &=&(-1)^{N_1} { \prod_{j=1}^{N_2} A(v_j)
 \prod_{j=1}^{N_3} A(\tilde z_j)} \ \
 \caA^+_{\vv\cup\uu}[{Q^-_{\thth^{(12)} }\over Q^+_{\thth^{(12)}}}]\,
 , \la{first factor} \\
 \langle  \ww ,\tilde  \zz\rangle_{\thth^{(3)}}\ \ \ 
& =&(-1)^{N_3}  \prod_{j=1}^{N_3} A(\tilde z_j)
 \ \ \caA^+_\ww[ {Q^+_{\thth^{(23)} }\over Q^-_{\thth^{(23)}}}] 
 .
\la{secondfactor}
 \ee
 Proof: Using the properties of the $\caA$-functional (Appendix
 \ref{appendix:caA}), we transform
  \be \caA^+_{\vv\cup\tilde \zz\cup\uu}[ {Q^-_{\thth^{(1)} }\over
  Q^+_{\thth^{(1)}}}] &=&\caA^+_{\vv\cup\tilde \zz\cup\uu}[
  {Q^-_{\thth^{(13)} }\over Q^+_{\thth^{(13)}}}{Q^-_{\thth^{(12)}
  }\over Q^+_{\thth^{(12)}}}]= \caA^+_{\vv\cup\uu}[{Q^-_{\thth^{(12)}
  }\over Q^+_{\thth^{(12)}}}], \\
 \caA^+_{\ww\cup\tilde \zz}[ {Q^-_{\thth^{(3)} }\over Q^+_{\thth^{(3)}}}]&=&
 \caA^+_{\ww\cup\tilde \zz}[ {Q^-_{\thth^{(13)} }\over Q^+_{\thth^{(13)}}}
{Q^-_{\thth^{(23)} }\over Q^+_{\thth^{(23)}}}]
=
 \caA^+_\ww[ {Q^-_{\thth^{(23)} }\over Q^+_{\thth^{(23)}}}]
.
 \ee
Ignoring the factors that  are pure phases, we find for  the cubic vertex 
\be \la{cubicsu3scp} \<\uu, \vv,\ww\>^\su3 =
\caA^+_{\vv\cup\uu}[{Q^-_{\thth^{(12)} }\over Q^+_{\thth^{(12)}}}] \
\caA^+_\ww[ {Q^-_{\thth^{(23)} }\over Q^+_{\thth^{(23)}}}]\, .  \ee
The second factor has been evaluated in a different ways in
\cite{Omar} and \cite{SL}, where it was used that it equal to a
partial domain wall partition function of the six-vertex model.
  
 The norm $\< \uu, \uu \> $ is most easily computed by taking the
 expression for the inner product $\<\vv, \uu\>$, Eq.  \re{tilSLA}, in
 the limit $\vv\to\uu$.
  
  \section{The $su(3)$ structure constant in terms
of the $\caA$-functional}
\la{sect:dilog.su3}

A generic Bethe state $ |\uu\rangle$ in an $su(3)$ sector is
characterized by the rapidities $\uu= \{\uu_1, \uu_2\}$ and the
inhomogeneity parameters $\thth$ associated with the momentum-carrying
node (1), where
\be \uu_1=\{u _{1,j}, \dots, u _{1, N}\} , \quad \uu_2=\{u _{2,1},
\dots, u _{ 2,M} \} \, , \quad \thth = \{ \th_1,\dots, \th_L\} .  \ee
The rapidities satisfy the nested Bethe wave functions for the $su(3)$
R-matrix given by \re{Rmatrix}:
\be \la{BAE3} \prod_{l=1}^L {u _{1, j} - \th_l+ \hf i \over u _{1, j}
-\th_l -\hf i } & =& - \prod_{n=1}^{N} {u _{1,j} - u _{1,n} + i \over
u _{1,j} - u _{1, n} -i }\ \prod_{m=1}^{M} {u _{1,j} - u _{2, m} - \hf
i \over u _{1,j} - u _{2,m} +\hf i } \no \\
1& =& - \prod_{m=1}^{M} {u _{2,j} - u _{2,m} + i \over u _{2,j} - u
_{2,m} -i }\ \prod_{n=1}^{N} {u _{2,j} - u _{ 1,n} - \hf i \over u
_{2,j} - u _{ 1,n} +\hf i } \, .  \ee

The Bethe equations \re{BAE3} follow from the requirement that that
the $su(3)$ $T$-matrix in the fundamental representation,

\be
T(u) = e^{ip^1_\uu(u)}+
e^{ip^2_\uu(u)}
+e^{ip^3_\uu(u)}\, ,
\ee
has vanishing residues at the Bethe roots.  For given distribution of
the roots $\uu_1$ and $\uu_2$, the pseudomomenta $p_\uu^i(z)$ are
determined modulo $2\pi$ by\footnote{Here we used the conventions of
Eqs.  \re{defQ} and \re{shifts}.  }
\be
\begin{aligned}
e^{i p^1_\uu}= {Q_{\uu_1}^{++}\over Q_{\uu_1}} {Q_\thth^{-}\over
Q_\thth^{+}} ,\quad e^{i p^2_\uu}= {Q_{\uu_1}^{--}\over Q_{\uu_1}} {
Q_{\uu_2}^{+} \over Q_{\uu_2}^{-} } , \quad e^{i p^3_\uu}&=
{Q_{\uu_2}^{---}\over Q_{\uu_2}^{-} } \, , \,
\end{aligned}
\ee
  see e.g. \cite{GV-complete1l}.  In terms of the three pseudomomenta,
  the Bethe equations \re{BAE} read
\be
\la{BAEquasi}
\begin{aligned}
e^{ip^1_\uu(z) -ip^2_\uu(z)}&= -1 \quad \text{if} \ z\in \uu_1; \\
e^{ip^2_\uu(z) -ip^3_\uu(z)}&= -1 \quad \text{if} \
z- i/2 \in \uu_2\, .
\end{aligned}
\ee

 It is convenient to introduce the functions $P^1_\uu(z)$ and
 $P^1_\uu(z)$, associated with the two nodes of the Dynkin graph of
 $su(3)$, and related to the quasimomenta $p^i_\uu(z)$, $i=1,2,3$, by
 \be P^1_\uu(z) = p^1_\uu(z) - p^2_\uu(z) , \quad P^2_\uu(z) =
 p^2_\uu(z+i/2) - p^3_\uu(z+i/2).  \ee
 In terms of these functions, which we will also call pseudomomenta,
 the Bethe equations take the more standard form
   \be
e^{iP^a_\uu(z)}=-1, \quad \text{if} \ z\in \uu_a\quad (a=1,2).
  \ee
  The functions $P_1$ and $P_2$ can be expressed in terms of the
  $su(3)$ Cartan matrix $ \{ M_{ab}\} = \(^{\ \ 2\ \ -1}_{-1\ \ \ \
  2}\)$
as
  \be
  e^{iP^a_\uu} =
  \(
 {Q_\thth ^-\over Q_\thth^+} \)^{ \d_{a, 1} }
\prod_{b=1,2} {
 Q_{\uu_b} ^{[M_{ab}]}
 \over
 \  Q_{\uu_b}^{[-M_{ab}]}}\, , \quad a=1,2. 
 \ee
Let us stress that the values of the local conserved charges are
determined only by the level-1 roots $\uu_1$.  The duality
transformations change the   level-2 roots $\uu_2$, but leave
invariant the   level-1 roots $\uu_1$, which carry the physical
information \cite{GV-complete1l}.

\medskip

\noindent{\bf The norm of an on-shell Bethe state} \ The squared norm
of an on-shell Bethe state has been computed for the case of $su(3)$
by Reshetikhin\footnote{A conjecture for $su(n)$ is proposed by EGSV
in \cite{EGSV}.} \cite{reshet-SU3} and is expressed as the determinant
of the matrix of the derivatives of the two
quasi-momenta:\footnote{Here it is assumed that the set of the Bethe
roots is symmetric under complex conjugation.}
   \be
   \<\uu, \uu \> = c_\uu\ \det\[ \partial_{u_{a,j} } P^b_{ \uu}
   (u_{b, k}) \] , \ee
where the determinant is with respect to the double indices $A=\{a,
j\}$ and $B=\{b, k\}$.  The normalizationn factor $c_\uu$ 
is given by (\ref{cu}).  The matrix of the derivatives of the two
quasimomenta is explicitly
 \be \la{ddCY} 
 \partial_{u_{a,j} } P^b_{ \uu} (u_{b, k}) &=&
 t_{ab}(u_{a,j} -u_{b, k}) +t_{ab}(-u_{a,j} +u_{b, k}) + i\,
 \d_{a,b}\d_{j,k} {\p P^{a}_{ \uu}(z)\over \p {z}}\Bigg|_{z=u_{a,j}}\,
 , \ee
  where \be t_{ab}(u) = {1\over u} - {1\over u+ {i\over 2} M_{ab} }.
  \ee
Instead of taking the derivatives, we will compute the norm as the
limit of the determinant depending on two sets of rapidities, $\uu$
and $\vv$, which has the limit \re{ddCY} when $\vv\to\uu$.  We define
the $(N+M)\times (N+M)$ square matrix $\O_{ab} (u_j, v_k)$, with \be
\la{defOa3} \O_{ab}(u, v)& =& t_{ab}(u - v) - e^{i P^a_\uu (v)} \
t_{ab}( -u+v)\, .  \ee
 The expression for the norm, which we are going to evaluate in the
 classical limit, is
  \be \la{normedet} \<\uu,\uu \> = c_\uu \lim_{v_{a,j}\to u_{a,j}}
  \det\[ \O_{ab} (u_{a,j} , v_{b, k}) \] .  \ee

\subsection{The
inner  product $\langle \uu, \vv \rangle$ in the  limit $\uu_2\to\infty$ }

Unlike the $su(2)$ case, the inner product of an on-shell Bethe state
with an of-shell Bethe state is not generically a determinant.
Determinant representations exist in some particular cases
\cite{joaocaetano-scpr, Wheeler-SU3, 2012arXiv1207.0956B}.  We will
use the determinant expression obtained by Wheeler \cite{Wheeler-SU3},
when the rapidities of the second type of magnons of the Bethe
eigenstate are sent to infinity.  We assume that $M$ is odd; then one
can send to infinity the $\uu_2$ roots one by one.  As a result the
second level Bethe equations become trivial and the first level Bethe
equations take the same form as for $su(2)$.  The inner product
$\<\uu,\vv\>^\su3 _\thth$ factorizes into two $su(2)$ inner products
\cite{Wheeler-SU3}
\be \la{WheelerA} 
\lim_{\uu_2\to\infty} \langle \vv, \uu\rangle^\su3
&=& {\det _{jk} \([v_{2, j}]^{k-1} - [v_{2, j}+i]^{k-1} {
Q_{\vv_1}^-(v_{2, j} ) \over Q_{\vv_1}^+(v_{2, j}) } \) } \no \\
 &\times& {\det_{ij}\( t(u_{1, j} -v_{1, k})\ - {Q_\thth^-(v_{1,k})
 \over Q^+_\thth(v_{1, k})}\ { Q_{\uu_1}^{++}(v_{1,k} ) \over
 Q_{\uu_1}^{--}(v_{1,k} )} \, t(-u_{1,j} + v_{1,k}) \)} \no \\
&\times& {1 \over
\Delta[\vv_1]\  \Delta[\vv_2]\ \Delta[\uu_1]}
\times \prod_{j,k} (u_{1,j}-v_{1,k} + i)
\no
\\
&=& \< \uu_1,\vv_1\>_\thth^{ su(2)}\ \< \bf{\infty}, \vv_2\>_{\vv_1}^{
su(2)}.  \ee
Using \re{PDWFS}, we write \re{WheelerA} in the form
 \be\la{su2caA}
 \begin{aligned}
\lim_{\uu_2\to\infty} \langle \vv, \uu\rangle_\thth
&=\caA^+_{\uu_1\cup\vv_1}[ {Q_\thth^-\over Q_\thth^+}]\, \
\caA^+_{\vv_2} [ {Q^-_{\vv_1}\over Q^+_{\vv_1}}] \, .
\end{aligned}
\ee

\medskip

\noindent{\bf The $su(3)$ structure constant in terms of the
$\caA$-functional.  } Combining \re{firstfactor}, \re{secondfactor},
\re{WheelerA} in Eq.  \re{cubicsu3}, we get
   \be \la{numeratorsu3} \<\uu,\vv,\ww\>
   =\caA_{\ww_1}^+[{Q_{\thth^{(13)}}^-\over Q_{\thth^{(13)}}^+}] \
   \caA^+_{\uu_1\cup\vv_1}[ {Q_{\thth^{(12)}}^-\over
   Q_{\thth^{(12)}}^+}]\, \ \caA^+_{\vv_2} [ {Q^-_{\vv_1}\over
   Q^+_{\vv_1}}] \, .  \ee

\section{The semi-classical limit of the $su(3)$ 3-point function}
\la{sect:semi-classical}

 \subsection{The Sutherland limit} 
The classical , or thermodynamical limit is attained for long spin
chains ($L\gg 1$) with macroscopically many excitations $N, M\sim L$,
and in the low energy regime $(\d E\sim 1/L)$ \cite{Beisert:2003xu,
Kazakov:2004qf, Beisert:2005di}.  Such spin chains correspond to
``heavy'' operators, which are traces of products of many SYM fields.
In this limit the roots scale as $u_{a, j}\sim L$.  In the condensed
matter literature the classical limit has been studied by Sutherland
\cite{PhysRevLett.74.816} and by Dhar and Shastry
\cite{PhysRevLett.85.2813}, and is known as Sutherland scaling limit.
In the classical limit the roots are organized in several macroscopic
strings, which condense into cuts in the complex rapidity plane.  The
three quasimomenta $p_1,p_2,p_3$ become the three branches of the same
meromorphic function.  The three sheets of the corresponding Riemann
surface are joined among themselves along the cuts defined by the long
Bethe strings.
In the classical limit, the Bethe state is characterised by the
resolvents
\be G_{\uu_1}(z)= \p_z \log Q_{\uu_1}(z), \quad G_{\uu_2}(z)=\p_z\log
Q_{\uu_2}(z), \ee
as well as the resolvent for the inhomogeneities
\be
  G_\thth(u) = \p_u \log Q_\thth (u).
  \ee
The two resolvents, $G_{\uu_1}$ and $G_{\uu_2}$, can be expressed in
terms of the three quasimomenta $p^1_\uu, p^2_\uu$ and $p^3_\uu$,
which become the three branches of a single meromorphic function on
the tri-foliated Riemann surface,
\be
\la{boundcdsa}
\begin{aligned}
p_\uu^1 &=   G_{\uu_1} - G_\thth \ \  (\text{mod}\ 2\pi),
\\
p^2_\uu  &=  G_{\uu_2} - G_{\uu_1}  \quad (\text{mod}\ 2\pi),
\\
p^3_\uu &= - G_{\uu_2} \quad (\text{mod}\ 2\pi).
\end{aligned}
\ee
or
\be  
\la{boundcdsb}
\begin{aligned}
P_\uu^1  &= 2 G_{\uu_1}- G_{\uu_2} - G_\thth \  (\text{mod}\ 2\pi),
\\
P^2_\uu &= 2 G_{\uu_2} - G_{\uu_1}  \quad (\text{mod}\ 2\pi).
\end{aligned}
\ee

Let $\CC^\a_{ij}$ be the cuts joining the $i$-th and the $j$-th
sheets.  Then the Bethe equations \re{BAEquasi} become boundary
conditions on these cuts, depending on the mode numbers $n^\a_{ij}$:
 \be
 \la{boundconda}
 \begin{aligned}
  2\pi n _{12}^\a &=\sla p_1 - \sla p_2 , \qquad z\in \CC^\a_{12}\, ,
  \\
  2\pi n _{23}^\a &=\sla p_2 - \sla p_3 , \qquad  z\in \CC^\a_{23}\, ,
\end{aligned}
 \ee
where $\sla p$ denotes the half-sum of the values of the function $p$
on both sides of the cut.

   \subsection{Stacks}  
In addition, there is the possibility of configurations called stacks
(bound states of rapidities associated with different nodes
\cite{Takahashi-BS}), which represent pairs of roots belonging to the
nodes 1 and 2 and at distance $O(1) $ from each other
\cite{Beisert:2004ag,Beisert:2005di,GV-complete1l}.  We can have
macroscopic strings of stacks, which in the classical limit become two
cuts that merge into one cut.  Since the roots that form the string of
stacks belong to two different nodes, they correspond to a cut type
1-2 and a cut type 2-3, where we understand that the cut of type
$i$-$j$ joins the $i$-th and the $j$-th sheets of the Riemann surface.
The result of merging of the two cuts is a cut of the type 1-3.
Therefore, in order to have a description of the generic Bethe state
in the classical limit, we must assume also the existence of cuts of
type 1-3.  The boundary condition on these cuts is obtained by taking
the limit of \re{boundconda} and has the form
 \be
 \la{boundcondb}
 \begin{aligned}
 2\pi n _{13}^\a &=\sla p_1 - \sla p_3 , \qquad z\in \CC^\a_{13}\, .
\end{aligned}
 \ee
The bosonic duality transformations \cite{GV-complete1l} in the
classical limit corresponds simply to  the exchange of the Riemann
sheets 2 and 3.

\subsection{The semi-classical norm}  
The determinant \re{normedet} can be computed in the classical limit
under the assumption that there are only 1-2 and 2-3 type cuts, which
are separated at macroscopic distance $\sim L$.  With this assumption,
the off diagonal elements of $\O(u, v)\sim (u-v)^{-2} \sim 1/L^2$, and
the only matrix elements of order one are those in a strip of width
$\sim 1/\sqrt{L}$ along the diagonal.  As a consequence, the
non-diagonal blocks do not contribute in the classical limit and the
determinant is simply the product of the determinants of the diagonal
blocks,
      \be
      \la{normedetdiag}
      \begin{aligned}
 \<\uu, \uu \>_\su3 &\simeq \det\[ \O_{11} (u_{1,j} , u_{1, k}) \]
 \
  \det\[  \O_{22} (u_{2,j} , u_{2, k}) \] 
 \\
 &= \< \uu_1 , \uu_1\> ^{su(2)}\ \< \uu_2, \uu_2\>^{su(2)} .
\end{aligned}
  \ee
Let us evaluate the norm assuming that there there are no cuts
relating the first and the third sheet of the Riemann surface.  We can
use the expression for the classical limit of the norm in the $su(2)$
sector (Appendix \re{appendix:caA}):
\be
\la{scprviaA}
\begin{aligned}
\< \uu, \vv\>_{\thth}= \log\caA ^+ _{\uu \cup \vv}
[{Q_\thth^-\over Q_\thth^+}]& = \oint\limits _{C_\uu\cup C_\vv}
{dz\over 2\pi}\ \Li[f(z) \ e^{ i G_\uu(z) +i G_\vv(z) - iG_\thth(z)}]+
o(\log L) ,
\end{aligned}
\ee
with 
\be
 G_\uu(z) = \p_z\log Q_\uu(z),\quad  G_\thth(z) = \p_z \log
Q_\thth(z).
\ee
The norm of the classical Bethe state is then\footnote{We conjecture 
that in the most general case, when some of the roots can form bound
states (``stacks"), this logarithm of the norm is given by
 \bigskip
 \be
 \la{clnorm}
 \begin{aligned}
   \log \< \uu|\uu\> &= \sum_{\a<\b} \oint_{\CC_{\a\b}}{dz\over 2\pi}
   \ \Li \big( e^{ i p^\a_\uu(z) - i p^\b_\uu(z) } \big) ,
 \end{aligned}
\ee
where $\CC_{ij}$ ($i,j=1,2,3)$ denote the contour (or contours)
surrounding the cuts between the $i$-th and the $j$-th sheets.  }
\medskip
\be
\la{scprno13}
 \log \< \uu|\uu\> = \oint_{\CC_{\uu_1}}{dz\over 2\pi} \
\Li \big( e^{ 2 iG_{\uu_1} (z) - iG_{\uu_2}(z) - iG_\thth(z) } \big) +
\oint_{\CC_{\uu_2}}{dz\over 2\pi} \ \Li \big( e^{ 2i G_{\uu_2} (z) -
iG_{\uu_1}(z)} \big) .
\ee
\medskip
 \vfill
\subsection{ Semi-classical limit of the structure constant} 
 Substituting \re{scprviaA} in \re{numeratorsu3} we find for the
 structure constant in the classical limit
\be
\begin{aligned}
\log C_{123}^{(0)}&= \oint_{\CC_{\uu_1\cup\vv_1}}{dz\over 2\pi} \ \Li
\big( e^{ i G_{\uu_1}(z) +iG_{\vv_1}(z) - iG_\thth^{(12)}(z) }\big) \\
      + \oint_{\CC_{ \vv_2}}{dz\over 2\pi} &\ \Li \big( e^{i
      G_{\vv_2}(z) - iG_{\vv_1}(z)}\big) + \oint_{\CC_{
      \ww_1}}{dz\over 2\pi} \ \Li \big( e^{i G_{\ww_1}(z) -
      iG_{\thth^{(13)}}(z)}\big) \\
     & - \hf \log \<\uu|\uu\> -\hf\log\<\vv |\vv\> -\hf \log\<
     \ww|\ww\>.
\end{aligned}
\ee
The last line is evaluated in the classical limit according to
\re{scprno13}.

\section{Conclusions and outlook}
\la{sect:conclusions}

We have analyzed the tree-level 3-point functions of single-trace
operators of the planar $\CN=4$ SYM theory in the $su(3)$ sector.
Each of the three operators is an eigenstate of the dilatation
operator, and it is characterized by a set of charges (angular
momenta) and a set of rapidities.  We have classified the possible
configurations of Wick contractions and given the general expression
of the 3-point functions in terms of rapidities associated to each
operator.  This expression, obtained using the {\it tailoring}
technique of EGSV \cite{EGSV}, is not adapted for taking the classical
limit.  In some particular situation, when one of the operators
belongs to an $su(2)$ sector, we are able to express the 3-point
function using the alternative method of {\it freezing} proposed in
\cite{Omar}.  By further specializing the second group of rapidities
corresponding to one of the operators, we can use a result of
\cite{Wheeler-SU3} to express the scalar products as determinants.
Finally, the semi-classical limit of the determinants can be taken
using the results from \cite{3pf-prl,SL}.  The simplest classical
operator from the $su(3)$ sector are the three-spin solution obtained
by C. Kristjansen \cite{Charlotte-3spin04ei}.

There are two obvious directions to explore.  First, one can try to
evaluate the general sum over partitions in \re{c123}
quasi-classically.  One can either try to perform a quasiclssical
evaluation of the Korepin sum over partitions for the scalar product
of two off-shell Bethe states, in the spirit of \cite {GSV}, or refine
the coherent state approximation
\cite{Kruczenski:2004aa,Escobedo:2011xw,BOH}.  Another direction is to
explore the non-compact sector of the theory.  There are several
recent papers which are relevant for that, \cite{2012arXiv1203.4246J,
2012arXiv1203.1913C, 2012arXiv1203.1617D, 2012arXiv1208.0841S,
2012arXiv1207.5489G, Caetano:2012ac, ConfRT, 2012arXiv1212.6563K,
Plefka:2012rd, ConfRT, 2012arXiv1207.3112E}.

  \section*{Acknowledgments}
We thank J. Caetano, N. Gromov, P. Vieira and M. Wheeler
 for useful discussions.
Part of this work has been supported by Institut Henry Poincar\'e, 
by the Australian Research
Council and the European Union Seventh Framework Programme
[FP7-People-2010-IRSES] under grant agreement N$^o$ 269217.

\appendix

\section{ The nested coordinate Bethe Ansatz}
\begin{figure}[h!]
\begin{center}
\includegraphics{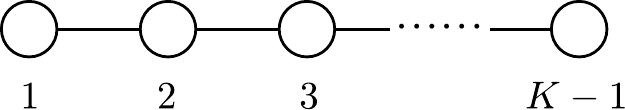}
\caption{The Dynkin diagram of $\mathfrak{su}(K)$ algebra.}
\label{Dynkin}
\end{center}
\end{figure}

\la{appendix:nesting} 
%\la{appendix:nested}
As an introduction to the `tailoring' procedure in the context 
of $su(3)$ 3-point function, this appendix is a brief introduction 
to the nested coordinate Bethe anstaz of the $su(3)$ spin chain.  
Let us consider an $su(K)$ spin chain of length $L$.  The treatment
follows closely \cite{escobedo2012integrability}.  The hamiltonian
reads
\begin{align}\label{hamiltonian}
H=\frac{\lambda}{8\pi^2}\sum_{n=1}^L(\mathbf{I}_{n,n+1}-\mathbf{P}_{n,n+1})\;.
\end{align}
At each site of the spin chain, there is a spin with $K$ different
polarizations.  The Hilbert space of the spin chain is
$\mathcal{H}=(\mathbb{C}^K)^{\otimes L}$.  In the Hamiltonian
(\ref{hamiltonian}) $\mathbf{I}_{n,n+1}$ is the identity operator in
the space $\mathbb{C}^K_n\otimes\mathbb{C}^K_{n+1}$ and
$\mathbf{P}_{n,n+1}$ is the permutation operator in
$\mathbb{C}^K_n\otimes\mathbb{C}^K_{n+1}$.  The Dynkin diagrams of the
Lie algebra give a convenient way to label the excitations.  For the
$\mathfrak{su}(K)$ algebra, the Dynkin diagram is given by
Fig.(\ref{Dynkin}).

Following Bethe, one  looks for the eigenstates of the Hamiltonian
(\ref{hamiltonian}) in  the  form
\begin{align}\label{ket}
|\Psi\rangle=\sum_{\text{positions}}\psi_{{\bf n}_1, \dots, {\bf
n}_{K-1}}(\mathbf{u}_1,\cdots,\mathbf{u}_{K-1})\,
|\mathbf{n}_1,\cdots,\mathbf{n}_{K-1}\rangle\;,
\end{align}
where $\mathbf{u}_a$ is the set of rapidities of node $a$ and
$\mathbf{n}_a$ labels the positions of the excitation of node $a$
$(a=1,\cdots,K-1)$.  The summation is taken over all possible
positions of excitations.  The nested ket state
$|\mathbf{n}_1,\cdots,\mathbf{n}_{K-1}\rangle$ is constructed in the
following steps:
\begin{enumerate}
\item Start with an initial value of length $L$; \item Create $N_1$
excitations of node 1 at positions $\mathbf{n}_1$.  The $N_1$
excitations form a reduced inhomogeneous spin chain of length $N_1$;
\item Create $N_2$ excitations of node 2 at positions $\mathbf{n}_2$
in the reduced spin chain.  One should have $N_2<N_1$.  The
excitations of node 2 again form a reduced inhomogeneous spin chain of
length $N_2$; \item Repeat the above steps for all the $K-1$ nodes.
\end{enumerate}

\begin{figure}[h!]
\begin{center}
\includegraphics[scale=0.7]{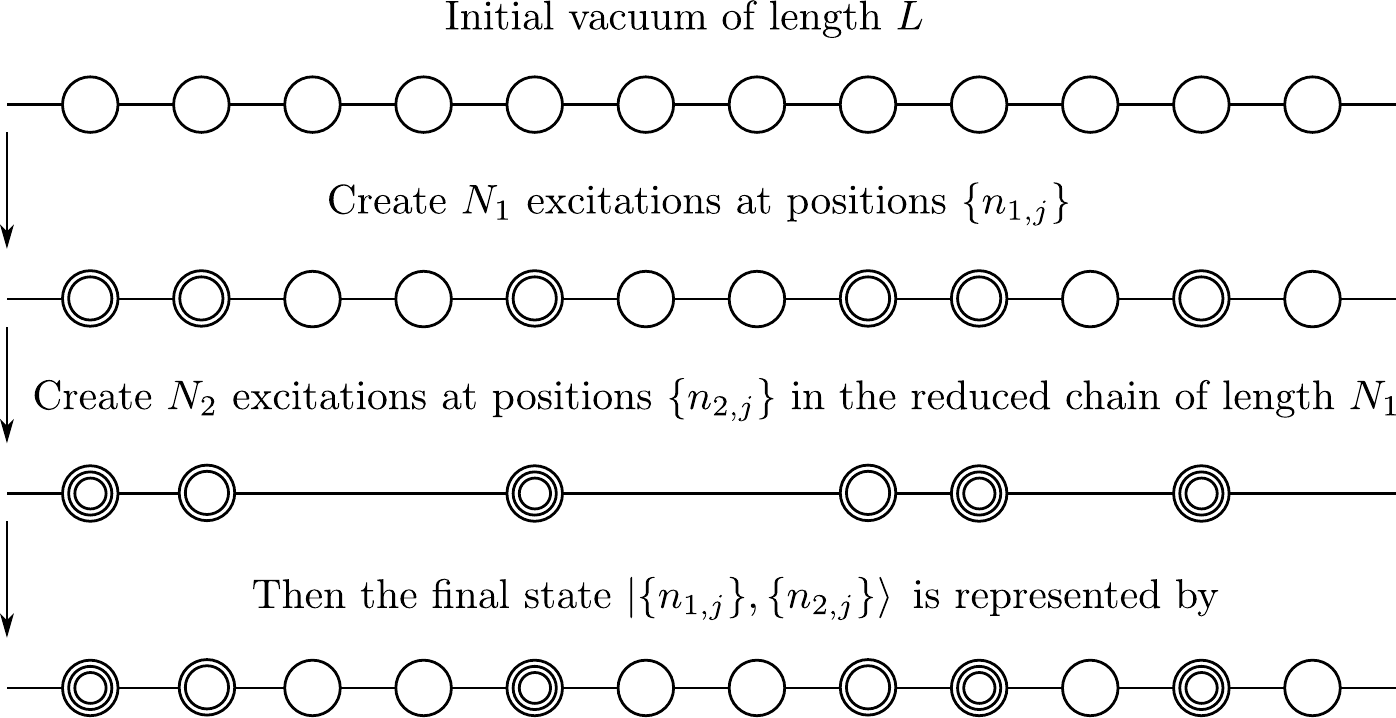}
\caption{\small
Nesting procedure for $su(3)$.  In our example, $K=3$,
$L=12$, $\{n_{1,j}\}=\{1,2,5,8,9,11\}$ and
$\{n_{2,j}\}=\{1,3,4,5,6\}$.  The final result reads
$|2,1,0,0,2,0,0,1,2,0,2,0\rangle$, where $0$ denotes the vacuum state,
$1$ denotes the first excited state and $2$ denotes the second excited
state.  In our ket notation, it is written as
$|\{1,2,5,8,9,11\},\{1,3,4,5,6\}\rangle$.  }
\label{nestket}
\end{center}
\end{figure}

The procedure for the $su(3)$ case is explained in
Fig.(\ref{nestket}).  We would like to stress the important feature
that the excitations of node $a$ should only be created from the
excitations of node $a-1$ (the reduced spin chain).  Therefore we have
$N_a\le N_{a-1}$.  The positions should obey
\begin{align*}
&1\le n_{1,1}<\cdots<n_{1,N_1}\le L,&\\
&1\le n_{a,1}<\cdots<n_{a,N_a}\le N_{a-1},\quad\text{for } 2\le a\le
r\;.&
\end{align*}
The beauty of this method is that at each step, the operation is
simple and the same: to create one kind of excitation in a reduced
spin chain of length $N_a$.  Here we see the \textsl{nested} feature
of this method.\par The wave function
$\psi(\mathbf{u}_1,\cdots,\mathbf{u}_{K-1})$ is constructed by nested
Bethe ansatz.  It is given in terms of a series of wave functions
$\psi_a$, $a=2,\cdots,K-1$.The Bethe wave function reads
\begin{eqnarray}\label{wave}
\psi(\mathbf{u}_1,\cdots,\mathbf{u}_{K-1})=\sum_{\mathrm{P}_1}A_1(\mathrm{P}_1)
\prod_{j=1}^{N_1}\left(\frac{u_{1,\mathrm{P}_{1,j}}+
\frac{i}{2}}{u_{1,\mathrm{P}_{1,j}}-\frac{i}{2}}\right)\psi_2(\mathrm{P}_1)\;,
\end{eqnarray}
where the wave functions $\psi_a$, $a=2,\cdots,K-1$ are given by
\begin{eqnarray}
\psi_a(\mathrm{P}_{a-1})=\sum_{\mathrm{P}_a}A_a(\mathrm{P}_a)
\prod_{j=1}^{N_a}\prod_{k=1}^{n_{a,j}}\frac{\left(u_{a,\mathrm{P}_{a,j}}-u_{a-1,
\mathrm{P}_{a-1,k}}-M_{a-1,a}\frac{i}{2}\right)^{\delta_{k\ne
n_{a,j}}}}{u_{a,\mathrm{P}_{a,j}}-u_{a-1,\mathrm{P}_{a-1,k}}
+M_{a-1,a}\frac{i}{2}}\psi_{a+1}(\mathrm{P}_a).
\end{eqnarray}
with $\mathrm{P}_a$ the permutation $(P_{a,1},\ldots, P_{a,N_a})$ of
$(1,\cdots,N_a)$.  We define $\psi_{K}(\mathrm{P}_{K-1})=1$.
$M_{a,b}$ is the Cartan matrix of the Lie algebra, for
$\mathfrak{su}(K)$ being
\begin{eqnarray}
M_{a,b}=2\delta_{a,b}-\delta_{a-1,b}-\delta_{a+1,b}\;.
\end{eqnarray}
Our choice of normalization is
\begin{align}
A_a(1,2\cdots,N_a)=1
\end{align}
and the coefficients $A_a$ obey the relation
\begin{align}
\frac{A_a(\cdots,i,j,\cdots)}{A_a(\cdots,j,i,\cdots)}=S_2(u_{a,i},u_{a,j})\;.
\end{align}
Here we used the following definition, 
\begin{align}
\label{s12}
S_\s(u_{a,i},u_{b,j})=
\frac{u_{a,i}-u_{b,j}+\frac{i}{2}\s}{u_{a,i}-u_{b,j}-\frac{i}{2}\s}\;,
\qquad \s=1,2\;.
\end{align}
In order that (\ref{ket}) is an eigenstate of the spin chain
Hamiltonian, the rapidities should satisfy the Bethe ansatz equations
:
\begin{align}\label{nBAE}
\left(\frac{u_{a,j}+V_a\frac{i}{2}}{u_{a,j}-V_a\frac{i}{2}}\right)^L
=\prod_{b=1}^r\prod_{k=1\atop
(a,j)\ne(b,k)}^{N_b}\frac{u_{a,j}-u_{b,k}+\frac{i}{2}M_{a,b}}{u_{a,j}-u_{b,k}
-\frac{i}{2}M_{a,b}}\;,
\end{align}
where $V_a$ are the Dynkin labels.  We consider the fundamental
representation in this paper, where $V_a=\delta_{a,1}$.

\section{The $su(3)$ tailoring prescription}
\la{appendix:tailoring}

We consider the operators in $su(3)$ sector with definite one-loop
anomalous dimensions.  In the spin chain language, these operators 
are represented by the Bethe eigenstates of the $su(3)$ spin chain.  
One first write the Bethe state as the entangled state of two subchain
states. This operation is called ``cutting''. After cutting operation, 
each subchain state also takes the form of Bethe states.
In order to perform Wick contraction, one needs to ``flip'' one of the
subchains.  Flipping is an operation that takes a ket state into the
corresponding bra state with the same wave function.  One can flip
either the left or the right subchain.  In this paper, we always flip
the right subchain.  The last step is to calculate the scalar product
of Bethe states, this is called the ``gluing'' operation.  

\begin{figure}[h!]
\begin{center}
\includegraphics[scale=0.7]{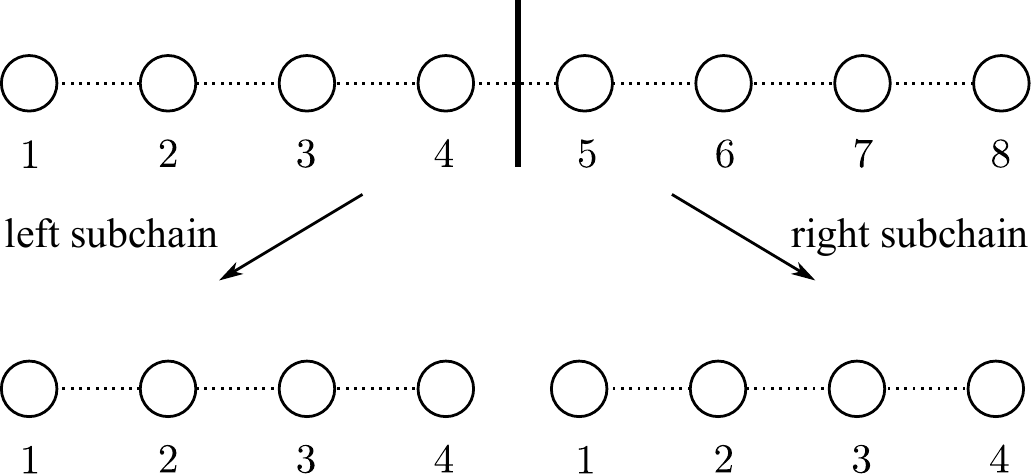}
\caption{\small The cutting process for a spin chain with $L=8$.  We take
$l=4$ and $L-l=4$.  The sites in the right subchain are originally
labeled by $5,6,7,8$ while after cutting they are labeled by
$1,2,3,4$.}
\label{cut}
\end{center}
\end{figure}

Consider a generic $su(2)$ Bethe state $|\mathbf{u}\rangle$ of a spin
chain with length $L$.  We define the first $l$ sites from the left to
be the \textsl{left} subchain and the rest $L-l$ sites to be the
\textsl{right} subchain.  $|\mathbf{u}\rangle$ can be written as an
entangled state of the subchains
\begin{align}
|\mathbf{u}\rangle=\sum_{k=0}^{\text{min}\{N,l\}}\sum_{1\le
n_1<\cdots<n_k\le l}\ \sum_{l\le n_{k+1}<\cdots\le
L}\psi(\mathbf{u})|n_1,\cdots,n_k\rangle_{\,l}
\otimes|n_{k+1}-l,\cdots,n_N-l\rangle_{\,r}
\end{align}
where $k$ is the number of magnons in the left subchain.  Note that
one needs to re-label the positions of the magnons in the right
subchain, see Fig.(\ref{cut}).
Since we have two subchains, the magnons can either be in the left
subchain or in the right subchain.  After cutting a Bethe eigenstate,
the resulting two subchains states still take the form of Bethe state.
Hence the two subchain states have their own Bethe wave functions
$\psi_l(\mathbf{u}')$ and $\psi_r(\mathbf{u}'')$, where $\mathbf{u'}$
and $\mathbf{u''}$ is a partition of $\mathbf{u}$,
\begin{align}
\mathbf{u}'\cup\mathbf{u}''=\mathbf{u},\quad\mathbf{u}'\cap\mathbf{u}''=\emptyset\;.
\end{align}
In general,
$\psi(\mathbf{u})=H(\mathbf{u}',\mathbf{u}'')\;\psi_l(\mathbf{u}')\,
\psi_r(\mathbf{u}'')$ where $H(\mathbf{u}',\mathbf{u}'')$ is a
partition-dependent factor which shall be called $H$-factor from now
on.  Formally, the cutting of a Bethe state can be written as
\begin{align}    
|\mathbf{u}\rangle=\sum_{\alpha}H(\mathbf{u}',\mathbf{u}'')\,|\mathbf{u}'\rangle\otimes|\mathbf{u}'' \rangle\;.
\end{align}
Usually the expression for the $H$-factor is long.  In order to make
the expression more compact, we introduce some short-hand notations.
Given a function $F(x,y)$ and two sets of variables $\mathbf{u}$,
$\mathbf{v}$, we define
\begin{align}
F(\mathbf{u},\mathbf{v})\equiv\prod_{u_i\in\mathbf{u},
\;v_j\in\mathbf{v}}F(u_i,v_j),\quad
F^>(\mathbf{u},\mathbf{v})\equiv\prod_{i>j\atop
u_i\in\mathbf{u},\;v_j\in\mathbf{v}}F(u_i,v_j)\;.
\end{align}
For a constant $c$, we define
\begin{align}
F(\mathbf{u},c)=\prod_{u_i\in\mathbf{u}}F(u_i,c),\quad
F(c,\mathbf{v})=\prod_{v_i\in\mathbf{v}}F(c,v_i)\;.
\end{align}
With the notations from \re{s12} the $H$-factor for the $su(2)$ spin
chain is given by
\begin{align}
H(\mathbf{u}',\mathbf{u}'')=S_1 (\mathbf{u}'',0)^l \; S_2^>
(\mathbf{u}',\mathbf{u}'')\;.
\end{align}
The cutting operation can be generalized to $su(K)$ Bethe state.  Let
us denote the nested Bethe state by $|{\mathbf{u}}\rangle$ where
$\mathbf{u}=(\mathbf{u}_1,\cdots,\mathbf{u}_{K-1})$.  We have
\begin{align}
|{\mathbf{u}}\rangle=\sum_{{\mathbf{u}}'}H({\mathbf{u}}',
{\mathbf{u}}'')\;|{\mathbf{u}}'\rangle\otimes|{\mathbf{u}}''\rangle
\end{align}
with the $H$-factor
\begin{align}
H({\mathbf{u}}',{\mathbf{u}}'')=\prod_{n=1}^{K-1}S_1
(\mathbf{u}''_n,\mathbf{u}'_{n-1})\;S_2^>
(\mathbf{u}'_n,\mathbf{u}''_n)
\end{align}
where $\mathbf{u}'_0$ is defined as $ \mathbf{u}'_0=\{0^l\} $ with $l$
the length of left subchain.  In order to perform Wick contraction, we
need to ``flip'' the right subchain from a ket state into a bra state.
The flipping operation is different from Hermitian conjugate.  Given a
state
\begin{align}
|\psi\rangle=e^{i\theta}|XZXZZ\rangle
\end{align}
the Hermitian conjugate and flipping (denoted by superscript
$\mathcal{F}$) lead to
\begin{align*}
(|\psi\rangle)^\dagger&=e^{-i\theta}\langle XZXZZ|\\
(|\psi\rangle)^{\mathcal{F}}&=e^{+i\theta}\langle
\bar{Z}\bar{Z}\bar{X}\bar{Z}\bar{X}|\;.
\end{align*}
For an $su(2)$ Bethe state $|\mathbf{u}\rangle$, the flipped state is
proportional to the hermitian conjugate of the Bethe state
$|\mathbf{u}^*\rangle$
\begin{align}
(|\mathbf{u}\rangle)^{\mathcal{F}}=\langle\mathbf{u}^*|F(\mathbf{u})
\end{align}
where $\mathbf{u}^*$ is the complex conjugate of $\mathbf{u}$ and we
call the proportionality $F(\mathbf{u})$ the $F$-factor.  For an
$su(2)$ Bethe state, the $F$-factor reads
\begin{align}
F(\mathbf{u})=S_1 (\mathbf{u},0)^{L+1}\;S_2^> (\mathbf{u},\mathbf{u})
\end{align}
where $L$ is the length of the spin chain.  For an $su(K)$ Bethe state
$|\vec{\mathbf{u}}\rangle$, the $F$-factor is given by
\begin{align}
F({\mathbf{u}})=\prod_{n=1}^{K-1}S_1
(\mathbf{u}_n,\mathbf{u}_{n-1})S_2^> (\mathbf{u}_n,\mathbf{u}_n)
\end{align}
where $\mathbf{u}_0$ is defined by $ \mathbf{u}_0=\{0^{L+1}\} $.  From
now on, by ``tailor'' (denoted by $\mathcal{T}$) a (nested) Bethe
state, we mean first cut the state and then flip the right subchain
state.  We define the product of the corresponding $H$-factor and the
$F$-factor to be the $\mathrm{H}_F$ factor.  For an $su(K)$ Bethe
state $|\mathbf{u}_1,\cdots,\mathbf{u}_{K-1}\rangle$
\begin{align}
(|{\mathbf{u}}\rangle)^{\mathcal{T}}=\sum_{{\mathbf{u}}'}
\mathrm{H}_F^\mathbf{u}\;
|{\mathbf{u}}'\rangle\otimes\langle{\mathbf{u}}''^*|
\end{align}
where
\begin{align}
{\mathbf{u}}'\cup{\mathbf{u}}''={\mathbf{u}},
\quad {\mathbf{u}}'\cap{\mathbf{u}}''=\emptyset
\end{align}
and the $H_F$-factor reads
\begin{align}
\mathrm{H}_F^\mathbf{u}=\prod_{n=1}^{K-1}S_1
(\mathbf{u}''_n,\mathbf{u}_{n-1})S_2^>
(\mathbf{u}_n,\mathbf{u}''_n)\;.
\end{align}

\section{The functionals $\caA^\pm$}
\la{appendix:caA}

 \subsection{Definition}
 For any set of points $\uu= \{u_j \}_{j =1}^N$ in the complex plane
 and for any complex function $f(z)$, we define the pair of
 functionals $\caA^\pm _\uu[f] $, which are completely symmetric
 polynomials of degree $N$ of the variables $f(u_1), \dots , f(u_N)$.
 The functional $\caA^\pm _\uu$ is defined as a sum of monomials
 labeled by all possible partitions of the set $\uu$ into two disjoint
 subsets $\uu'$ and $\uu''$, with $\uu' \cup\uu'' = \uu$,
\be \la{ExpA} \caA_{\uu}^\pm [f]&=& \sum_{\uu'\cup \uu''=\uu }\ \
\prod_{u' \in\uu' } [- f(u' )] \prod _{ u''\in \uu'' }{ u' - u'' \pm i
\over u' -u'' } .  \ee
The expansion \re{ExpA} was thoroughly studied by Gromov, Sever and
Vieira \cite{GSV}.  The expansion \re{ExpA} is summed up by the
following operator expression,
 \be \caA^\pm _\uu[f] = \hat \caA^\pm _\uu[f]\cdot 1\, , \ee
 where the different operator  $\hat\caA^\pm [\uu]$ is defined as

\be \la{defCA}
% \encadremath{
 \begin{aligned}
 \hat \caA^\pm _\uu[f] &\defeq { 1 \over \Delta_{\uu} } \prod_{j=1}^N
 \(1 - f(u_j ) \, e^{\pm i \p/\p u_j } \) \Delta_{\uu}, \qquad
 \Delta_{\uu} =\sum _{j<k} (u_j-u_k).
 \end{aligned}
%}
 \ee
   The operator functional $\hat\caA^\pm [\uu]$ is formally obtained
   from the c-functional $\caA^\pm [\uu]$ as
 \be \hat \caA^\pm _\uu[f]= \caA^\pm _\uu[f \, e^{i\p}]\, .  \ \ee

 \subsection{Properties}

It was found in \cite{GSV} that for constant function $f(u) = \k$, the
expansion \re{ExpA} does not depend on the positions of the rapidities
$\uu$ and the functional $\caA^\pm _\uu[f]$ is given in this case by
\be \la{triviale} \caA^\pm _{\uu}[\k ]= (1-\k)^N
 .  \ee
 The functional $\caA^\pm[f]$ have other remarkable properties
 \cite{3pf-prl, SL}:

 \bigskip \noindent 1) {\it Determinant formula}, Eq.
 \re{detformula}, which is obtained by substituting $\Delta_\uu=
 \det_{jk} (u_j^{k-1})$ in the definition \re{defCA}.

 \bigskip \noindent 2) {\it Functional relations between $ \caA^+ $
 and $ \caA^- $:}
   \be \la{funceqa} \caA^\mp _{\uu}[f]&=& \caA^\pm _{\uu}[1/f] \
   \prod_{j=1 }^N [- f(u_j )] \, , \la{funceqb} \\
    \caA^\mp _{\uu}[f]&=& \caA^\pm_\uu \big[ -
   {Q^{\mp\mp}_\uu\over Q^{\pm\pm }_\uu} \, f\big] \, .
   \ee
  \bigskip \noindent 3) {\it Reduction formula:}
\be \la{AuzA} \caA^\pm _{\uu\cup\zz}[f {Q_\zz\over Q^{ \pm\pm}_\zz}] =
\caA^\pm_\uu[f]\, .  \ee

 \bigskip \noindent 4) {\it Factorization property:}
\be\la{factorhat} \hat \caA_ {\uu\cup \vv}^\pm[f] =\hat \caA_ \uu^\pm[
{Q_\vv^{\pm\pm}\over Q_\vv} \, f] \cdot \hat \caA_ \vv^\pm[
{Q^{\pm\pm}_\uu\over Q_\uu} \, f]\, .  \ee
 Aplying $N$ times the factorisation property, one obtains the
 representation
\be \la{defCAht} \hat \caA^\pm _\uu[ f ] &=& \prod_{j =1}^N \(1 - \,
E^\pm_{j} f(u_j )\ e^{i \p/\p u_j}\) ,\qquad
%  \\
   E^\pm_j
   %& \defeq & \prod_{k (\ne j)} {u_j - u_k \pm i\over u_j-u_k}
   = \underset{u\to u_j}{ \text{Res}} {Q^{\pm\pm}_\uu(u)\over Q_\uu}
   .
 \ee

  \bigskip \noindent
  {\bf Relation to the Slavnov determinant.} The
  determinant of the kernel $\O(u,v)$ defined as
\be
\O_f (u,v) = t(u-v) - {Q^{++}_\uu\over Q^{--}_\uu}(u) \,  f(v) \, t(-u+v) ,
\ee
where the set $\uu$ satisfies the ``on-shell'' condition
\be \quad {Q^{++}_\uu(u_j) \over Q^{--}_\uu(u_j)} \ f(u_j)=-1 \quad
\text{for} \ u_j\in\uu, \ee
is evaluated as \cite{sz}
\be
\la{PDWFS}
{\det_{jk} \O_{f} (u_j, v_k) \over \det_{jk} {1\over u_j-v_k+i}}
= \caA^+_{\uu\cup \vv} [f].
\ee

\subsection{Classical limit. }
In the classical limit, the Bethe roots condensate in one or several
disjoint cuts.  Let $\CC_k$ be a contour encircling the $k$-th cut
anticloskwise and leaving outside all other singularities of $f$ and
$G_\uu$.  The filling fraction of the $k$-th cut is \be \a_k= {1\over
2\pi L} \oint \limits_{\CC_k} G_\uu(u) du.  \ee
We consider the limit $L\to\infty$ with all $\a_k$ finite.
 Then  the leading, linear in $L$,  term of $\log \caA^\pm $
 is given     by the contour integral    
 %   %
   \be \la{scalingA} 
   \log\caA^\pm_\uu [ f] & \simeq&    \pm \oint
   \limits_{ \CC_\uu} \frac{du}{2\pi } \ \text{Li}_2\big(f(u)\, e^{\pm
   iG_\uu(u)}\big) \, , \quad \CC_\uu = \cup_{k=1}^n\CC_k.  \ee
 While there is not yet a rigorous proof of this formula, it has
 passed a number of analytical and numerical checks.  A heuristic
 derivation of \re{scalingA} for $f(u) =\( {u-i/2\over u+i/2}\)^L$,
 was presented in \cite{GSV}.  When $f(u) = \k$, it was shown in
 \cite{GSV} that the quasiclassical formula \re{scalingA} gives the
 exact answer \re{triviale}.  Moreover, the integral \re{scalingA}
 satisfies the functional equation \re{funceqa} thanks to the
 functional equation for the dilogarithm,
\be \la{Lifunceq} \Li({1\over \o} )= - \Li(\o) - {\pi^2\over 6} -
{1\over 2} \log^2(-\o).  \ee
%

%%
%\bibliography{/Users/vani/Files/PAPERS/PAPERSLIBRARY/ABib}
%\bibliographystyle{/Users/vani/Files/PAPERS/PAPERSLIBRARY/utcaps}
%
 
 \providecommand{\href}[2]{#2}\begingroup\raggedright\endgroup

\end{document}